\documentclass[preprint]{elsarticle}

\usepackage[dvipsnames,table,xcdraw]{xcolor}
\usepackage{xspace}

\usepackage{soul}

\usepackage{tabularx}
\newcolumntype{Y}{>{\centering\arraybackslash}X}
\usepackage{enumitem}
\setlist{
  partopsep=0pt,
  topsep=2pt,
  itemsep=0pt,
  parsep=2pt
}

\usepackage{wrapfig}
\usepackage{multicol}
\usepackage{multirow}
\usepackage{booktabs}
\usepackage{float}
\usepackage{amsthm}
\usepackage{amsmath}
\usepackage{nccmath}
\usepackage{graphicx}
\usepackage{amsfonts}
\usepackage{caption}
\usepackage{mdframed}
\usepackage{listings}
\usepackage{xcolor}
\usepackage{courier}
\usepackage{caption}[2015/09/20]

\lstdefinelanguage{JavaScript}{
  keywords={typeof, new, true, false, catch, function, return, null, catch, switch, var, if, in, while, do, else, case, break, import, export, {*}, as, from, this},
  keywordstyle=\color{blue}\bfseries,
  ndkeywords={class, export, boolean, throw, implements, import, this},
  ndkeywordstyle=\color{darkgray}\bfseries,
  identifierstyle=\color{black},
  sensitive=false,
  comment=[l]{//},
  morecomment=[s]{/*}{*/},
  commentstyle=\color{PineGreen}\ttfamily,
  stringstyle=\color{purple}\ttfamily,
  morestring=[b]',
  morestring=[b]"
}

\usepackage[colorlinks=true, allcolors=blue]{hyperref}

\urldef{\exportsURL}\url{https://nodejs.org/api/modules.html\#modules\_exports}
\urldef{\definePropertyURL}\url{https://developer.mozilla.org/en-US/docs/Web/JavaScript/Reference/Global\_Objects/Object/defineProperty}

\theoremstyle{plain} 
\newtheorem{answer}{AQ}

\begin{document}

\begin{frontmatter}

\title{Automated Refactoring of Legacy JavaScript Code to ES6 Modules}

\author[aueb]{Katerina Paltoglou}
\ead{paltoglouk@aueb.gr}
\author[aueb]{Vassilis E. Zafeiris\corref{cor1}}
\ead{bzafiris@aueb.gr}
\author[aueb]{N. A. Diamantidis}
\ead{nad@aueb.gr}
\author[aueb]{E. A. Giakoumakis}
\ead{mgia@aueb.gr}

\address[aueb]{Department of Informatics, Athens University of Economics and Business,\\ 76 Patission Str., Athens 104 34, Greece}

\cortext[cor1]{Corresponding author. Address: Patission 76, 104 34, Athens, Greece. Tel: +30 210 8203 501}

\journal{Journal of Systems and Software}

\begin{abstract}
The JavaScript language did not specify, until ECMAScript 6 (ES6),
native features for streamlining encapsulation and modularity.
Developer community filled the gap
with a proliferation of design patterns and module formats,
with impact on code reusability, 
portability and complexity of build configurations.
This work studies the automated refactoring of legacy ES5 code
to ES6 modules with fine-grained reuse of module contents
through the \emph{named import/export} language constructs.
The focus is on reducing the coupling of refactored modules 
through destructuring exported module objects to fine-grained module features
and enhancing module dependencies by leveraging the ES6 syntax.
We employ static analysis to construct
a model of a JavaScript project, 
the Module Dependence Graph (MDG),
that represents modules and their dependencies.
On the basis of MDG we specify
the refactoring procedure
for module migration to ES6. 
A prototype implementation 
has been empirically evaluated on 19 open source projects.
Results highlight the relevance of the refactoring
with a developer intent for fine-grained reuse.
The analysis of refactored code shows
an increase in the number of reusable elements per project
and {\color{black}{reduction}} in the coupling of refactored modules.
The soundness of the refactoring is empirically validated
through code inspection and execution of projects' test suites.  
\end{abstract}


\end{frontmatter}

\newcommand{\meta}[1]{
  \noindent
  \begin{tabularx}{\textwidth}{c X} \hline
    {\color{JungleGreen}\Large\textsc{note}} & #1 \\\hline
  \end{tabularx}
}

\newcommand{\changeStart}{\color{black}}
\newcommand{\changeEnd}{\color{black}}
\newcommand{\change}[1]{{\color{black}#1}}

\newcommand{\code}[1]{\lstinline!#1!}
\newcommand{\codew}[1]{\texttt{#1}}
\newcommand{\valpct}[2]{#1 (#2\%)}
\newcommand{\pct}[1]{#1\%}
\newcommand{\avg}[2]{#1 (#2)}
\newcommand{\colhead}[2]{\multicolumn{1}{|p{#1}|}{\raggedright\textbf{#2}}}
\newcommand{\Ca}[0]{$C_a$\xspace}
\newcommand{\Fin}[0]{$F_{in}$\xspace}
\newcommand{\Ce}[0]{$C_e$\xspace}
\newcommand{\Fout}[0]{$F_{out}$\xspace}
\newcommand{\I}[0]{$I$\xspace}
\newcommand{\avgFO}{$\overline{\text{FO}_i}$\xspace}
\newcommand{\avgFI}{$\overline{\text{FI}_i}$\xspace}
\newcommand{\avgMI}{$\overline{\text{I}_i}$\xspace}
\newcommand{\avgDI}{$\overline{\Delta\text{I}_i}$\xspace}
\newcommand{\FO}{$\text{FO}_i$\xspace}
\newcommand{\FI}{$\text{FI}_i$\xspace}
\newcommand{\MI}{I$_i$\xspace}

\newcommand{\templateMethodDP}{\texttt{TEMPLATE METHOD}}
\newcommand{\nullObjectDP}{\texttt{NULL OBJECT}}

\newcommand{\express}{\texttt{express}}

\newcommand{\javascript}{JavaScript\xspace}
\newcommand{\jscodeshift}{jscodeshift\xspace}
\newcommand{\jshint}{JSHint\xspace}
\newcommand{\tern}{Tern\xspace}

\newcommand{\eval}{\texttt{eval}}
\newcommand{\this}{\codew{this}}
\newcommand{\expressjs}{\texttt{express.js}}
\newcommand{\requestjs}{\texttt{request.js}}
\newcommand{\applicationjs}{\texttt{application.js}}
\newcommand{\routerindexjs}{\texttt{router/index.js}}
\newcommand{\queryjs}{\texttt{query.js}}
\newcommand{\initjs}{\texttt{init.js}}
\newcommand{\viewjs}{\texttt{view.js}}
\newcommand{\utilsjs}{\texttt{utils.js}}
\newcommand{\routerjs}{\texttt{router.js}}

\newcommand{\GluttonousSnake}{\texttt{GluttonousSnake}}
\newcommand{\astix}{\texttt{astix}}
\newcommand{\gameOfLife}{\texttt{game-of-life}}
\newcommand{\UltraTetris}{\texttt{UltraTetris}}

\newcommand{\dataJs}{\texttt{data.js}}

\newcommand{\tetrisjs}{\texttt{tetrisJS}}
\newcommand{\scorejs}{\texttt{Score.js}}
\newcommand{\tetrisgamejs}{\texttt{TetrisGame.js}}
\newcommand{\boardjs}{\texttt{Board.js}}
\newcommand{\appTetrisGame}{\texttt{app/TetrisGame}}
\newcommand{\appBoard}{\texttt{app/Board}}
\newcommand{\playing}{\texttt{PLAYING}}
\newcommand{\clearing}{\texttt{CLEARING}}
\newcommand{\domjs}{\texttt{dom.js}}
\newcommand{\dom}{\texttt{dom}}
\newcommand{\self}{\texttt{self}}
\newcommand{\lodashjs}{\texttt{lodash.js}}
\newcommand{\score}{\texttt{Score}}
\newcommand{\windowGame}{\texttt{window.game}}

\newcommand{\lodash}{\texttt{lodash}}
\newcommand{\microtpl}{\texttt{micro-tpl}}
\newcommand{\jugglingdb}{\texttt{jugglingdb}}
\newcommand{\goojs}{\texttt{goojs}}

\newcommand{\script}{\texttt{<script>}}

\newcommand{\hangmanHTML}{\texttt{hangman.html}}
\newcommand{\hangman}{\texttt{Hangman}}
\newcommand{\contentJs}{\texttt{content.js}}
\newcommand{\logicJs}{\texttt{logic.js}}
\newcommand{\manJs}{\texttt{man.js}}
\newcommand{\stopwatchJs}{\texttt{stopwatch.js}}
\newcommand{\hintJs}{\texttt{hint.js}}
\newcommand{\audiocontrolsJs}{\texttt{audiocontrols.js}}
\newcommand{\hintsleft}{\texttt{hintsleft}}
\newcommand{\hintsleftMutator}{\texttt{content\_set\_hintsleft}}

\newcommand{\memoryJs}{\texttt{memory.js}}
\newcommand{\storeJs}{\texttt{store.js}}
\newcommand{\session}{\texttt{express-session}}
\newcommand{\memoryStore}{\texttt{MemoryStore}}
\newcommand{\exports}{\texttt{exports}}
\newcommand{\moduleexports}{\texttt{module.exports}}
\newcommand{\require}{\texttt{require}}
\newcommand{\exStore}{\texttt{store\_Store}}
\newcommand{\store}{\texttt{Store}}
\newcommand{\sessionVar}{\texttt{Session}}
\newcommand{\sessionAlias}{\texttt{session\_Sessionjs}}
\newcommand{\sessionJs}{\texttt{session.js}}

\newcommand{\bodyJs}{\texttt{Body.js}}
\newcommand{\planckjs}{\texttt{planck.js}}
\newcommand{\Solver}{\texttt{Solver.js}}
\newcommand{\Settings}{\texttt{Settings.js}}
\newcommand{\mathjs}{\texttt{Math.js}}
\newcommand{\vecJs}{\texttt{Vec3.js}}
\newcommand{\VecTwo}{\texttt{Vec2.js}}

\newcommand{\define}{\texttt{define}}
\newcommand{\tetrisJS}{\texttt{tetrisJS.js}}
\newcommand{\ScoreJS}{\texttt{Score.js}}
\newcommand{\Score}{\texttt{Score}}
\newcommand{\requirejs}{\texttt{requirejs}}
\newcommand{\lodashEntry}{\texttt{lodash}}
\newcommand{\domEntry}{\texttt{dom}}
\newcommand{\lodashVar}{\texttt{\_}}
\newcommand{\domVar}{\texttt{dom}}
\newcommand{\commonJS}{\texttt{common.js}}
\newcommand{\jqueryJS}{\texttt{jquery.js}}
\newcommand{\eventsJS}{\texttt{events.js}}
\newcommand{\tetrisGameJS}{\texttt{TetrisGame.js}}
\newcommand{\common}{\texttt{common}}
\newcommand{\jquery}{\texttt{jquery}}
\newcommand{\events}{\texttt{events}}
\newcommand{\tetrisGame}{\texttt{app/TetrisGame}}
\newcommand{\game}{\texttt{game}}
\newcommand{\window}{\texttt{window}}
\newcommand{\globalObj}{\texttt{global}}
\newcommand{\undefinedVal}{\texttt{undefined}}
\newcommand{\globalGame}{\texttt{globalGame}}
\newcommand{\refext}[1]{~\texttt{\detokenize{#1}}}
\newcommand{\getGlobalGame}{\texttt{getGlobalGame}}
\newcommand{\setGlobalGame}{\texttt{setGlobalGame}}
\newcommand{\commonJSFile}{\texttt{common.js}}
\newcommand{\commonEntry}{\texttt{common}}

\newcommand{\jshintJs}{\texttt{jshint.js}}
\newcommand{\jshintRefJs}{\texttt{jshint\_refactored.js}}
\newcommand{\jsHintRefJs}{\texttt{jshint\_refactored.js}}
\newcommand{\JSHINTVar}{\texttt{JSHINT}}
\newcommand{\Lexer}{\texttt{Lexer}}
\newcommand{\lexJs}{\texttt{lex.js}}
\newcommand{\vars}{\texttt{vars}}
\newcommand{\varsJs}{\texttt{vars.js}}
\newcommand{\scopeManager}{\texttt{scopeManager}}
\newcommand{\scopeManagerJs}{\texttt{scope-manager.js}}

\newcommand{\umdJs}{\texttt{umd.js}}
\newcommand{\returnExports}{\texttt{returnExports}}
\newcommand{\bjs}{\texttt{b.js}}
\newcommand{\bentry}{\texttt{b}}
\newcommand{\bobject}{\texttt{b}}

\newcommand{\import}{\texttt{import}}
\newcommand{\export}{\texttt{export}}
\newcommand{\exportDef}{\texttt{export default}}

\newcommand{\req}{\texttt{req}}
\newcommand{\app}{\texttt{app}}
\newcommand{\proto}{\texttt{proto}}
\newcommand{\view}{\texttt{View}}
\newcommand{\query}{\texttt{query}}
\newcommand{\router}{\texttt{Router}}
\newcommand{\middleware}{\texttt{middleware}}
\newcommand{\compileTrust}{\texttt{compileTrust}}
\newcommand{\compileQueryParser}{\texttt{compileQueryParser}}
\newcommand{\compileETag}{\texttt{compileETag}}

\newcommand{\foobarJs}{\texttt{foobar.js}}
\newcommand{\testJs}{\texttt{test.js}}
\newcommand{\globalVar}{\texttt{globalVar}}
\newcommand{\globalVarBinding}{\texttt{globalVarBinding}}

\newcommand{\windowIndexing}{\texttt{window[<index>]}}
\newcommand{\globalIndexing}{\texttt{global[<index>]}}

\newcommand{\filename}{\texttt{__filename}}
\newcommand{\dirname}{\texttt{__dirname}}

\section{Introduction} \label{introduction}

JavaScript followed, 
in the last decade, the wide reach of the web platform 
and turned into a general-purpose programming language
for the mobile and server platform.
Despite the growing scope and size of JavaScript applications,
the language specification did not include, until recently,
native features for streamlining encapsulation and modularity~\cite{Wirfs2020}.
Syntactic constructs for modules and classes,
along with other features for writing complex applications
(e.g., promises, arrow functions)
were introduced in the ECMAScript 2015 (ES6) release~\cite{es6spec}.
 Since ES6 classes represent syntactic ``sugar'' over function constructors
and prototype-based inheritance,
they were implemented in browser engines soon after ES6 release.
On the other hand, native support for ES6 modules in popular browsers
was introduced after the first quarter of 2017~\cite{mdnJSRef},
while it is still an experimental feature
of the latest release of the Node.js runtime~\cite{nodejsEsm}.

ES6 modules support the development of complex applications
through encapsulation of top-level variable and function declarations 
in a JavaScript file in a new scope, the \emph{module} scope,
that differs from the global scope.
Thus, top-level declarations 
do not ``pollute'' the global scope,
preventing name collisions and enabling variable checking. 
Such incompatibilities in the execution model of ES6 code files
with respect to ES5 or earlier versions,
required non-trivial extensions to JavaScript engines
that needed time to be tested and released.

As \change{native} support for modularity 
lagged with respect to the needs of the industry,
the developer community filled the gap
with the proliferation of design patterns,
module formats and tools.
The encapsulation of state and behaviour
and the avoidance of global scope ``pollution''
were handled by code patterns  
(e.g. Immediately Invoked Function Expressions- IIFEs)
and module formats, 
mainly AMD~\cite{amdspec} and CommonJS~\cite{nodejsModules},
\change{which introduced} interfaces 
for declaring modules and their dependencies
in the browser and the server respectively.

Since ES6 modules are currently supported in most JavaScript runtimes, 
developers are confronted with the need
to migrate their legacy codebases
to the standard module format.
Besides simplification of code and build configurations,
the migration provides performance improvements,
as module loading and dependency management in ES6
are natively supported by the runtime,
rather than external libraries.
Moreover, special features of ES6 modules improve code maintainability:
(a) named imports/exports enforce name consistency within the codebase,
facilitate refactoring operations 
and allow for optimization of application size (tree-shaking) during deployment,
(b) named exports act as live immutable bindings 
that enhance safety by preventing errors relevant to cyclic dependencies~\cite{Rauschmayer15},
(c) eager resolution of the reachable dependencies of a module (e.g. the application entrypoint),
provides fail-fast program behaviour~\cite{Shore2004}.

In this paper, we propose a fully-automated method
for migration of an ES5 codebase to ES6 modules.
The method applies to code that is either non-modular
or uses the AMD/CommonJS module formats
and operates on the Module Dependence Graph (MDG),
a global model of the codebase.
The MDG provides a representation of the code 
in terms of modules, exported module features and module dependencies,
as inferred and optimized by the our method.
The proposed migration
represents a large and complex refactoring,
since it affects the entire code-base
and the transformation of each individual source file
requires analysis of all incoming and outgoing dependencies.
Besides handling the mapping among 
different module formats,
our method focuses on reducing 
the coupling of refactored modules through:
(a) selective destructuring of existing AMD/CommonJS module objects
to fine-grained module features, 
(b) establishment of the minimum required dependencies among modules 
by leveraging the ``named imports/exports'' ES6 language construct and
(c) encapsulation of global state and functions as cohesive module features.

\changeStart
As part of this work,
we explore the relevance of the proposed refactoring
and evaluate its impact on system modularity and code correctness
through an empirical study
that provides answers to the following research questions:  

\begin{enumerate}
    \item How prevalent is the presence of ES5 code, 
    modular or not, in real-world applications?
    \item Is there a developer intent 
    for independently reusable features within modules?
    \item What are the effects of the proposed refactoring
    to system modularity?
    \item Does the proposed source code transformation
    preserve the external behaviour of the analyzed system?
\end{enumerate}

Among the main findings of our empirical study
is the wide use of legacy module formats in open source projects
and the frequent presence of code patterns in legacy code
that reveal a need for individually reusable module features
within modules.
As concerning the effects on system modularity,
our method contributes to fine-grained reuse among system modules
and a reduction in module coupling.
Finally, the successful execution 
of the benchmark projects' test-suites on refactored code
provides empirical evidence on the soundness 
of the proposed source code transformation.

The rest of the paper is organised as follows. 
Section \ref{relatedwork} presents related work
and Section \ref{moduleSystemReview}, 
provides a brief review on modularity support in ES5
and compares it against the features of ES6. 
In Section \ref{transformationProcedure}, 
we specify our method for refactoring 
towards ES6 fine-grained modularity. 
Section \ref{evaluation}
presents the details and results
of an empirical evaluation of the method,
while Section \ref{threats}
discusses the threats to its validity.
Finally, Section~\ref{results} provides
a discussion of results and
the paper is concluded in Section \ref{Conclusions}.

\changeEnd
\section{Related Work} 
\label{relatedwork}

Refactoring involves reorganizing source code 
while preserving its external behaviour~\cite{Mens2004}. 
Fowler~\cite{Fowler99} introduced a series of basic refactorings
for handling common code smells in object-oriented code
and specialized them on the JavaScript language
in the second edition of his book~\cite{Fowler18}. 
Recent studies and 
systematic literature reviews focus on
the relationship between 
code smells and refactoring, 
especially refactoring towards 
mitigating architectural 
problems~\cite{Sousa2020, Soares2020}, and 
\change{their impact on 
software quality~\cite{Bavota2015,Lacerda2020,Fernandes2020,Agnihotri2020}}. 
Ongoing research expands 
the application of refactoring techniques
to address diverse requirements,
beyond design improvements in code 
or other software artifacts,
e.g. UML diagrams~\cite{Baqais2020}.
Specifically, several works focus on techniques for 
\changeStart
improving security~\cite{Abid2020} and  
maintainability metrics~\cite{Nasagh2020}, 
improving performance through 
removing unused features~\cite{Bruce2020}, 
reasoning about refactoring activities over 
long periods of time~\cite{Brito2020} 
as well as improving software  accessibility~\cite{Paiva2020} 
and achieving business-oriented goals~\cite{Ivers2020}.
\changeEnd

This work belongs to the research area
of refactoring methods and techniques that target the JavaScript language.
Relevant works in this area study the automation of refactorings
that handle the following concerns:   
(a) improvement of the code's runtime performance,
(b) improvement of software quality characteristics 
through generic refactorings or 
replacement of insecure language features and 
(c) migration of legacy code to newer language versions.
These works are outlined in the rest of this section.

\citet{Ying2013} leverage 
static analysis techniques for
refactoring AJAX applications
to shift from the XML to the JSON data format, 
in order to improve the runtime performance. 
Furthermore, static and dynamic analysis techniques 
are proposed towards automating
\changeStart 
dead code elimination in JavaScript applications
\cite{Obbink2018, Vazquez2018}. 
Feldthaus et al.~\cite{Feldthaus11,Feldthaus13}
\changeEnd
improve modularity through 
providing a framework for specifying generic refactorings 
based on pointer analysis. 
~\citet{Jensen12} propose a static analysis technique
for replacing \texttt{eval} with language features 
that enable reasoning about the program's behaviour.

Several approaches focus on 
refactoring JavaScript legacy code 
towards using language features provided in 
more recent versions of ECMAScript. 
The prevalence of callbacks in 
JavaScript applications 
and their equivalence to ES6 promises 
\changeStart
is studied in~\cite{Gallaba2015, Brodu2015, Gallaba2017}.
The identification of classes 
and their dependencies in legacy JavaScript
is studied 
in~\cite{Rostami16,SilvaSANER17}.
Extensions of these works 
provide migration rules for their refactoring 
towards using ES6 classes~\cite{SilvaJSEP17, SilvaICSR17}.
\citet{Paltoglou2018} propose 
a static analysis technique for 
refactoring non-modular ES5 web applications 
towards ES6 modules. 
\changeEnd

In this paper, we propose a method for automated refactoring 
of a code-base implemented in ES5 or earlier versions, 
towards fine-grained modularity through ES6 \emph{named imports/exports}.
Our work handles concerns related to code migration to a newer language version,
as well as to the improvement of its internal quality attributes. 
\changeStart
As compared to earlier work on refactoring legacy ES5 code
to the ES6 class syntax~\cite{SilvaJSEP17,SilvaICSR17},
our work is complementary and a basic requirement for its application.  
\changeEnd
The reason is that migration to ES6 modularity
is a prerequisite for further application of migrations
that incorporate modern language features to a code-base.
Concerning our previous work on this area~\cite{Paltoglou2018},
this paper contributes major extensions in terms of
(a) the scope of the refactoring, since it additionally handles AMD/CommonJS code-bases,
(b) the effect of the refactoring, 
as it changes module structure and enables 
effective usage of ES6 modularity constructs 
and (c) the evaluation methodology 
that highlights the effectiveness and practicality of the approach.  

\section{Modularity in JavaScript} \label{moduleSystemReview}

In this section, we review the evolution 
of modularity from ES5 to ES6.
We briefly present the 
CommonJS and AMD module formats
that are widely used in ES5
and their mostly adopted implementations~\cite{Rauschmayer2014},
followed by the ECMAScript Modules (ESM) 
feature standardized in ES6. 
\changeStart
Details on the projects containing 
the presented fragments are provided in Section~\ref{evaluation}.
\changeEnd
We focus on disparities concerning 
the encapsulation of module and global state,
the granularity of exported definitions 
and the evaluation order of module dependencies. 
Moreover, we refer to the imposed limitations with respect 
to the scope of applicability of the provided features.
In line with the overview of the features provided by ES6, 
we compare ES5 formats against ES6
and highlight their incompatibilities, 
which complicate automated migration of ES5 code to ES6.

\subsection{Modularity in ES5} \label{es5ModuleSystems}

Unlike static programming languages JavaScript did not provide
prior to ES6, a formal mechanism for 
preserving modularity and encapsulation \cite{Osmani08}. 
The community addressed this need  
through adopting design patterns, 
e.g. the Singleton pattern~\cite{Zakas12}, 
\changeStart
that inject data and functionality 
belonging to the global namespace 
into a single global variable.
\changeEnd

\begin{figure}[htb]
 \centering
 \captionsetup{justification=centering}
 \includegraphics[scale=0.7]{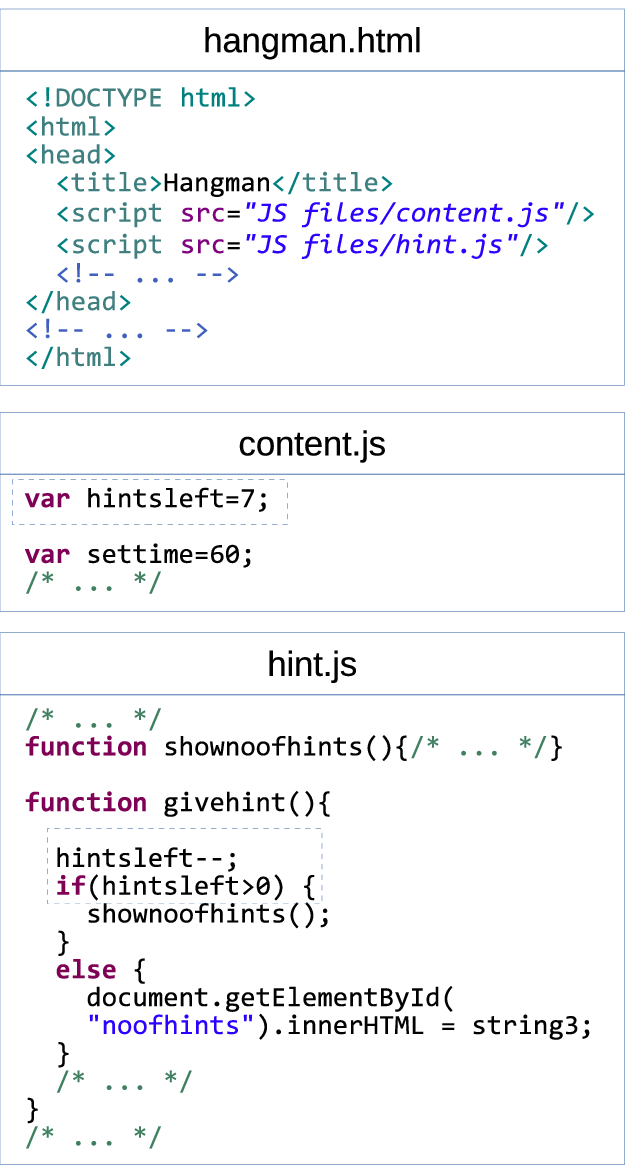}
 \caption{Declaration and access to global variables in ES5.}
 \label{fig:es5-js-snippets}
\end{figure}
 
Although each JavaScript file is meant to
bundle coherent functionality,
representing a conceptual module, 
the variables/functions declared in 
the file's top-level scope
are bound to the global object after the file is loaded.
Global variables are, also, introduced 
in terms of using non-declared variables 
(implied globals) \cite{Crockford08} 
in non-strict mode code \cite{Zakas12} 
and attaching properties to the global object.
Figure~\ref{fig:es5-js-snippets} 
depicts fragments from ES5 files of the \hangman{} project
that are loaded in page \hangmanHTML{}.
Upon loading, \contentJs{}
defines the global variable \hintsleft{} 
and makes it available to \hintJs{}.

The main problem arising from 
loading files with non-modular ES5 code is 
that it complicates program state management 
due to global variables. 
The presence of global variables in 
different program units 
hampers reasoning about the origin of their values, 
notably when these units are located in different files \cite{Zakas12}. 
Furthermore, it undermines maintainability,
since variables loaded from different files
increase the likelihood of name conflicts 
leading to runtime errors or unexpected behaviour \cite{Crockford08}. 
Finally, the lack of a module loading mechanism 
increases the application's complexity, 
since the module loading order 
must be manually specified.
The developer community handled these challenges
with the CommonJS and AMD module formats, 
that support the declaration of modules 
on the server and the browser, respectively~\cite{Rauschmayer15}.

\subsubsection{CommonJS} \label{commonJSSection}

\begin{figure}[htb]
    \centering
    \captionsetup{justification=centering}
    \includegraphics[scale=0.7]{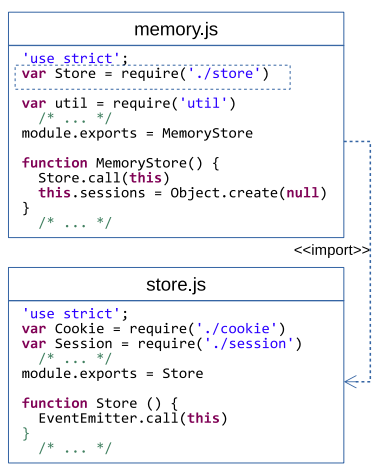}
    \caption{Module definition and loading using the CJS format.}
    \label{fig:commonJSExample}
   \end{figure}

The CommonJS (CJS) proposal specifies 
an interface for defining and loading modules
on the server \cite{nodejsModules}. 
A CJS module comprises 
a piece of reusable JavaScript code
which makes objects available to other modules \cite{Osmani08}. 
Figure~\ref{fig:commonJSExample} presents
code fragments from \session{} Node.js package.

\changeStart
The top-level variables/functions of a CJS module  
are bound to the module scope,
since CJS restricts the visibility
of top-level declarations~\cite{nodejsModules}.
Similarly to non-modular ES5, 
global variables are introduced in terms of implied globals  
in non-strict mode code~\cite{Zakas12,Crockford08},  
and global object properties.
CJS modules are synchronously loaded
through \require{}~\cite{Osmani08}, 
a function which returns the \emph{module object}, 
i.e. the object containing the variables/functions
that are visible beyond the scope
of their declaring module~\cite{Rauschmayer15}.
They behave as singleton objects,
since their instances are cached
at the time of their first load~\cite{nodejsModules}.
A variable/function defined in the module scope
becomes available to other modules
through binding to the \moduleexports{} object  
or its alias \exports{}
\footnote{https://nodejs.org/api/modules.html\#modules\_exports}.
CommonJS does not pose limitations
regarding the scope of applicability of the require/export features, 
since they can be used in the module's inner scopes 
(e.g. nested in conditional statements).
\changeEnd

As depicted in Figure~\ref{fig:commonJSExample}, 
\memoryJs{} imports 
the function \store{} exported from \storeJs{}. 
The function is defined 
at the time it is exported from \storeJs{} 
due to variable/function declaration hoisting \cite{Rauschmayer15},
i.e. a reference to \store{} is bound to 
the nearest scope in which it is defined.

\changeStart
Notice that CJS partially enhances encapsulation, 
since it does not prevent 
the introduction of implied global variables~\cite{Crockford08}. 
The global namespace ``pollution'' due to implied globals
\changeEnd
is mitigated through the \codew{strict mode} directive \cite{Zakas12}
that prohibits certain language features.
Finally, software portability cannot be guaranteed, 
since the format targets at defining modules 
for server-side tasks.
The use of design patterns 
(e.g Universal Module Definition - UMD~\cite{umdspec})
is needed in order to load CJS modules in the browser.
These patterns produce overhead, 
in terms of memory space and time, 
in order to resolve the execution environment
of the loaded module \cite{Osmani08}.

\subsubsection{Asynchronous Module Definition (AMD)} \label{amdSection}

The AMD proposal specifies an interface for 
declaring and loading modules on browsers~\cite{amdspec}. 
\change{Similarly to non-modular ES5,
AMD supports global variables
through either the declaration of top-level variables, 
the definition of implied globals in non-strict mode 
code~\cite{Zakas12,Crockford08} 
or property bindings to the global object}.

AMD modules are loaded 
either \emph{eagerly} or upon request
with the help of the features provided by 
the format or script loaders, 
e.g. RequireJS, respectively.
By eager loading we mean that
upon request for loading an AMD module,
all its dependencies are loaded
before the execution of the requested module's code. 
They are instantiated through the function \define{} 
provided by the format \cite{Zakas12}, 
which returns the module object. 
\changeStart
However, AMD does not provide a feature
for the actual loading of modules upon request,
and, thus, script loaders (e.g. RequireJS~\cite{requirejs}) 
are used for module loading and dependency management.
\changeEnd

\begin{figure}[htb]
 \centering
 \captionsetup{justification=centering}
 \includegraphics[scale=0.7]{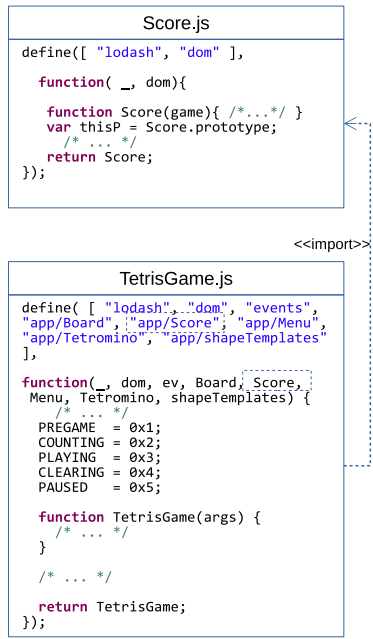}
 \caption{Module definition and loading using the AMD format.}
 \label{fig:AMDExample}
\end{figure}

Similarly to the CJS format, 
the features provided by the AMD format and 
the respective script loaders can be used in 
the module's inner scopes. 
Moreover, AMD modules are loaded in an asynchronous manner. 
The module loading and 
execution order cannot be precisely assumed, 
which can lead to the introduction of runtime 
errors\footnote{https://requirejs.org/docs/api.html\#data-main}. 
Finally, the problems encountered in defining AMD modules 
correspond to these encountered in 
defining CJS modules. 

\changeStart
Figure~\ref{fig:AMDExample} presents AMD module fragments 
from the \tetrisjs{} application, 
which are loaded with the RequireJS loader~\cite{requirejs}.
\changeEnd
The \tetrisgamejs{} module depends on \ScoreJS{}, 
which is referenced through 
the entry \codew{app/Score}  
in the dependency array provided in \define{}. 
The resolution of \ScoreJS{} 
imports its module object, 
i.e. the function \Score{}, 
in the scope of the module's definition function. 

\subsection{ES6 Modules} \label{esharmony}

The support for code modularity through 
the AMD/CJS formats
is currently replaced by 
native language features standardized in ECMAScript 6. 
These features enable defining 
platform-independent modules which 
can be loaded without the need for 
adopting design patterns (e.g. UMD~\cite{umdspec}).

The variables/functions defined 
in the top-level scope of an ES6 module 
are restricted to the module scope~\cite{Rauschmayer15}. 
Global variables are introduced in terms of 
global object properties, 
since ES6 modules enforce \textit{strict} mode 
which prohibits the introduction of 
implied global variables~\cite{Zakas12}. 

\begin{figure}[htb]
 \centering
 \captionsetup{justification=centering}
 \includegraphics[scale=0.7]{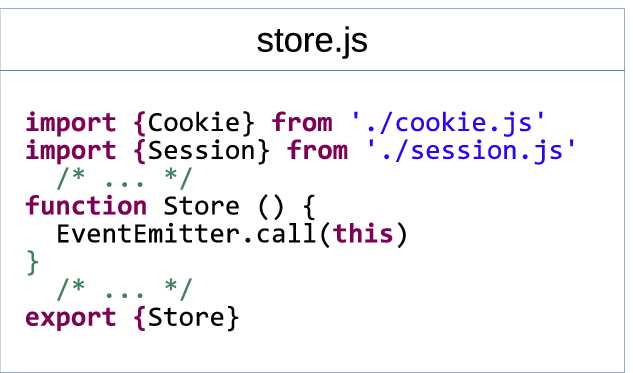}
 \caption{Module definition of Fig.~\ref{fig:commonJSExample} in ES6.}
 \label{fig:commonjs-vs-es6}
\end{figure}

ECMAScript 6 provides a declarative syntax for 
exporting and importing declarations 
through \export{} and \import{} statements.
Unlike AMD/CJS modules, 
which export a single \emph{module object}, 
i.e. a compound object with the module's exported declarations, 
an ES6 module can apply multiple strategies for managing dependencies. 
Specifically, it may use \emph{named exports}
to export multiple \emph{module features},
i.e. independent variable/function declarations,
to client modules.
Moreover, an ES6 module may selectively import
individual module features from another module
through static \textit{named} imports. 
Importing all module features as a compound object is,
also, supported through \textit{namespace} imports. 
Furthermore, each module may declare a specific feature
as a \emph{default export} 
that is imported in client modules 
through a \emph{default import} statement.
Finally, ES6 enables module loading
at module execution time through \textit{dynamic} imports.
Figure~\ref{fig:commonjs-vs-es6} presents 
the equivalent ES6 module of \storeJs{} 
(Figure~\ref{fig:commonJSExample}). 
The module declares the function \store{}, 
exported through a \emph{named export}, 
while it imports \sessionVar{} from \sessionJs{}
through a static \emph{named import}. 

Although ES6 standardized features 
for platform-independent modules,
their semantics pose a number of restrictions. 
Specifically, imported declarations introduce 
immutable bindings to 
the respective exported declarations 
in the module's scope~\cite{Rauschmayer15}. 
Moreover, ES6 exports and static imports 
are permitted only at 
the top-level scope of the module. 
The establishment of 
these statements in the module's inner scopes 
introduces evaluation errors. 

Compared to ES5 module formats,
ES6 modules support more effective elimination of unused code
through the use of \emph{named exports}
that simplify the identification of unused module features. 
Moreover, ES6 module dependencies 
are eagerly evaluated~\cite{es6spec},
before the actual execution of the module's code. 
Specifically, the module features 
that are imported in an ES6 module 
are resolved during the module's evaluation stage, 
in contrast to ES5 formats where 
they are resolved at module execution time~\cite{amdspec, nodejsModules}. 
This enables the identification of software defects 
relevant to non-declared imported features 
at the module's loading stage instead of 
its execution stage, thus providing fail-fast behaviour~\cite{Shore2004}. 
Finally, the ES6 module format is platform-independent,
enabling module reuse across different platforms~\cite{Grover2017}.
This, also, enhances testability~\cite{Zakas12} 
through mitigating the need for simulating 
the module's execution environment.

\section{Refactoring towards using ES6 modules} \label{transformationProcedure}

We propose a method for fully-automated migration
of legacy ES5 code to ES6 modules
that applies to ES5 projects that are non-modular
or employ the AMD/CJS formats
\change{as implemented by RequireJS~\cite{requirejs} and Node.js~\cite{nodejsModules}, respectively}. 
The method focuses on reducing the coupling of refactored modules, 
through destructuring existing module objects
to fine-grained module features
and optimization of module dependencies
on the basis of the minimum required module features.
Moreover, our method restricts the scope of global declarations, 
and confines them to the module scope as appropriate module features.

\changeStart

\begin{figure}[hbt]
   \centering
   \captionsetup{justification=centering}
   \includegraphics[scale=0.6]{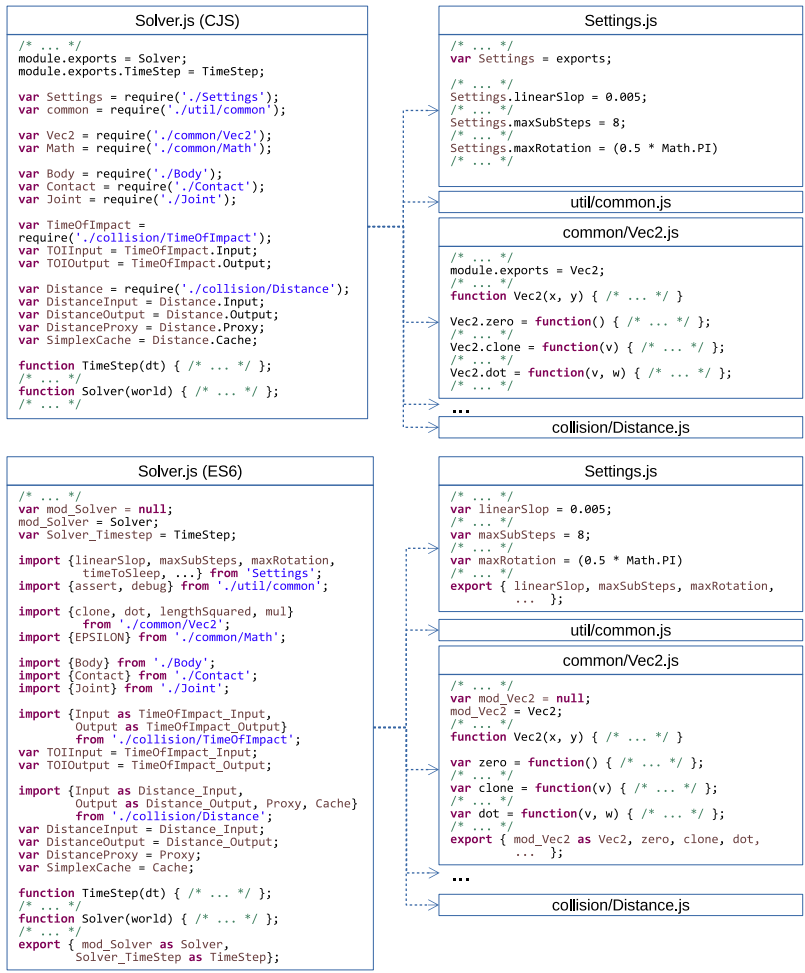}
   \caption{\change{Motivating example from the project \codew{planck.js}.}}
   \label{fig:motiv-example}
\end{figure}

Figure~\ref{fig:motiv-example} illustrates a motivating example,
in order to showcase the application and
the merits of the proposed refactoring.
It presents code from the \planckjs{} 2D physics engine
for game development and, specifically, 
code fragments from the \Solver{} module and its dependencies.
\Solver{} is a core module of \planckjs{}
and uses declarations from 9 modules of the same project,
e.g. it reads configuration properties from the \Settings{} module
and uses utility functions for vector calculations 
from the \VecTwo{} module.
The code is based on the CJS module format,
where the reusable declarations of a module
are exported as part of a single module object 
(top part of Figure~\ref{fig:motiv-example}).
For instance, 
the \Settings{} module exports 26 configuration properties,
as part of the \codew{Settings} object, 
\VecTwo{} exports 27 utility functions,
as bound properties to the \codew{Vec2} constructor function,
and \mathjs{} exports 8 functions and a constant
as part of the module object.
In CommonJS, each client of these modules, 
e.g., the \Solver{} module,
imports at once all their reusable features
through the respective module object.
Thus, the module clients depend on complex, coarse-grained objects
instead of the elementary features that they actually use.
In the case of \Solver{}
and imported modules \Settings{}, \VecTwo{} and \mathjs{},
the former depends on their entire public API,
although it uses 8, 4 and 1 of their exported features, respectively.

The bottom part of Figure~\ref{fig:motiv-example} 
presents an ES6 version of the code fragments
from \Solver{} and its dependencies,
after the application of the proposed refactoring.
Notice that the independently reusable features of each module,
e.g. configuration properties in \Settings{} or
utility functions in \VecTwo{},
are refactored to standalone declarations
and exported to client modules
through ES6 \emph{named export} statements.
The module structure is simplified,
while empty module objects are eliminated,
as is the case for the \codew{Settings} object.
Moreover, all declarations that a module requires from other modules  
are introduced to its scope 
through static \emph{named import} statements.
The \Solver{} module, now imports
the minimum required declarations from its dependencies.
\changeEnd

Fine-grained module design,
in terms of multiple independently reusable module features,
accompanied with fine-grained module dependencies
through named imports,
boosts refactorings, code optimizations and error tracing.
Specifically, it streamlines the renaming of module features
and their relocation among modules.
It, also, supports module bundling tools in effective tree-shaking, 
i.e., identification and elimination of unused code
from deployment artifacts~\cite{rollupjs, webpackTreeShaking}. 
Moreover, it enforces a static module structure which, in turn, 
paves the way for static variable checking
and type inference~\cite{Rauschmayer15}, 
enabling, thus, early error detection. 
Named imports, also, enable fail-fast program behaviour,
since missing imports are resolved at module evaluation stage
(probably at application startup),
rather than at a later execution stage~\cite{Shore2004}.

Besides \emph{named imports} or \emph{named exports},
the proposed method uses ES6 \emph{default import} statements  
for the migration of module dependencies on external libraries,
since they support backwards compatibility with ES5 module formats.
As concerning ES6 \emph{namespace} imports, 
i.e. import of all module features as a compound object,
they are not used by our method, 
since they introduce implicit dependencies among modules.
Specifically, in a namespace import,
the client module does not explicitly specify  
the module features that it depends upon.
Instead, it depends on all the features of the imported module,
including currently unused features
or features added in subsequent revisions.

In addition to the benefits provided by 
fine-grained modularity,
the proposed refactoring improves 
code portability and testability~\cite{Zakas12},
due to the migration of AMD/CJS modules to platform-independent modules 
and global variable encapsulation. 
The automation of the proposed refactoring 
is a challenging task 
due to incompatibilities of 
the ES6 module format against ES5 formats, such as
read-only bindings for imported module features, 
top-level declaration of static imports,
hoisting and eager loading of static module dependencies,
as opposed to on-demand loading in CommonJS etc.

In the rest of this section,
we specify our method for automated migration to ES6 modules
as two distinct processing stages:
(a) identification of refactoring candidates, 
i.e., code fragments that need to be refactored and
(b) application of the source code transformation.
\changeStart
The specification of our method, also, 
defines the refactoring preconditions
that preclude transformations
\changeEnd 
leading to erroneous code 
or influencing the analyzed program's external behaviour.


\subsection{Refactoring candidate identification procedure} \label{refactoringCandidateIdentificationProcedure}

The identification of refactoring candidates involves:
(a) the resolution of the structure of each module,
in terms of module features 
that can be individually exported and reused by other modules,
and (b) the resolution of fine-grained module dependencies,
on the basis of the minimum required module features 
that have to be imported by each module after refactoring.

The outcome of this procedure is a directed graph, 
the Module Dependence Graph (MDG),
that models the desired module structure after refactoring 
and supports the source code transformation procedure.
The MDG concept, initially introduced in our previous work~\cite{Paltoglou2018},
is redefined here to provide a unified representation of the module structure
in modular and non-modular JavaScript code. 
Specifically, it comprises a pair $MDG = (M, D)$,
where $M$ represents the set of graph nodes.
MDG nodes correspond to the ES6 modules 
that will be introduced after refactoring.
The set $M$ is initially populated
with one node for each JavaScript file of the analyzed project.
The set $D$ represents graph edges.
Each edge models a dependency 
between two modules of the refactored system.

\subsubsection{Resolution of module structure} \label{sect:resolve-mod-features}

This step involves the identification of 
the module features for each module in $M$.
Let $m_i\in M$ be a module 
and $n_i$ be its module name used in AMD/CJS 
or its module path in the case of non-modular ES5.
We define $m_i$ as a pair $m_i = (n_i, F_i)$,
where $F_i$ represents the set of module features of $m_i$.
Module features comprise 
(a) properties of the module object -- in AMD/CJS --
that can be extracted to module scope 
and get individually exported,
(b) the module object per se, 
in case that it is not stripped of all its content,
(c) global declarations 
that are assigned to modules in order to restrict their scope.

\paragraph{Module object destructuring}

The identification of object properties
that can be extracted as standalone module features
and their subsequent extraction
from the AMD/CJS module object 
is termed in this work as \emph{module object destructuring}.
The identification procedure involves
recovery of the module structure from its AST representation
and is based on the analysis of statements
that apply property bindings to the module object.

Property binding statements comprise assignments
whose left-hand side expression is an object property access
through dot/bracket notation, for instance \code{foo.bar = <expression>}.
\changeStart
Properties defined through bracket notation are excluded from further analysis,
as explained in~\ref{sect:preconditions}.
The \emph{bound properties} $B_i$ of module $m_i$, henceforth, 
refer to all property bindings on the module object $e_i$ with the dot notation.
\changeEnd
In case that the module object is instantiated
through object literal notation,
all properties appearing in the property-value pairs
are, also, appended to the set $B_i$.

The module object $e_i$ exported by module $m_i$
may be constructed within $m_i$ or imported from another module.
Its instantiation may be declarative,
through a function declaration or object literal notation,
or may involve the evaluation of an expression
(e.g. function invocation).
A module object instantiated as (a) an empty object/function
with properties attached at runtime,
or (b) through object literal notation,
is characterized by our method as a \emph{namespace} object.
The reason is that such objects
serve as containers for data/ function properties
that need to be exported by an AMD/CJS module.
Namespace objects would be stripped of all their properties
after refactoring, and, thus,
are eliminated from the set of module features.
Figure~\ref{fig:namespace-obj} presents
a module from \goojs{} project 
that follows the namespace object pattern.
The module exports mathematical constants and functions
and its original version is depicted in sub-figure (a).
Sub-figures (b) and (c) provide syntactic variants
that use an empty object or object literal notation.

\begin{figure}[htb]
  \begin{minipage}[t]{.5\textwidth}
  \begin{lstlisting}[language=JavaScript,
      captionpos=t,title={(a) Function object},
      basicstyle=\ttfamily\scriptsize,
      xleftmargin=2em, numbers = left]
  function MathUtils() {}
  
  MathUtils.DEG_TO_RAD 
             = Math.PI/180.0;
  /* ... */
  MathUtils.radFromDeg = 
    function(degrees){/* ... */};
  
  MathUtils.degFromRad = 
    function(radians){/* ... */};
  /* ... */
  module.exports = MathUtils;
  \end{lstlisting}
  \end{minipage}%
  \begin{minipage}[t]{.5\textwidth}
  \begin{lstlisting}[language=JavaScript,
      captionpos=t,title={(b) Empty object},
      basicstyle=\ttfamily\scriptsize,
      xleftmargin=2em, numbers = none]
  var MathUtils = {};
  
  MathUtils.DEG_TO_RAD 
             = Math.PI/180.0;
  /* ... */
  MathUtils.radFromDeg = 
    function(degrees){/* ... */};
  
  MathUtils.degFromRad = 
    function(radians){/* ... */};
  /* ... */
  module.exports = MathUtils;
  \end{lstlisting}
  \end{minipage}
  \begin{minipage}[t]{.3\textwidth}
  \begin{lstlisting}[language=JavaScript,
      captionpos=t,title={(c) Object literal},
      basicstyle=\ttfamily\scriptsize,
      xleftmargin=3em, numbers = left]
  var MathUtils = {
  
    DEG_TO_RAD: Math.PI/180.0,
    /* ... */,
    radFromDeg: function(degrees){/* ... */},
    degFromRad: function(radians){/* ... */},
    /* ... */
  };
  module.exports = MathUtils;
  \end{lstlisting}
  \end{minipage}
    \caption{Namespace object variants for \codew{MathUtils} module
  in \codew{goojs} project.}
    \label{fig:namespace-obj}
\end{figure}

Let $e_i$ be the module object of a module $m_i$
and $F_i$ the identified module features.
The module object destructuring algorithm computes the set $F_i$
for a module $m_i$ and can be summarized as following:
\begin{enumerate}[label=\emph{Step \arabic*.},leftmargin=1.2cm]
  \item Evaluate module object destructuring preconditions
  (see Section~\ref{sect:preconditions}).
  If any precondition fails then return $F_i = \{e_i\}$, 
  i.e., the module object cannot be destructured.
  \item Identify the set $B_i$ of bound properties to $e_i$,
  \item If $e_i$ is a \emph{namespace object} return $F_i = B_i$. 
  Otherwise, return $F_i = B_i \cup \{e_i\}$,
  i.e., the module object would be available as a module feature 
  after destructuring

\end{enumerate}

\paragraph{Resolution of global declarations}

Global declarations involve 
explicitly declared and implied global variables,
properties bound to platform specific global objects 
(e.g., \window{}, \globalObj{}) and
top-level declarations in non-modular ES5 source files.
We identify the global variables 
of a module $m_i$ through AST traversal and search for
(a) top-level \codew{var} statements (non-modular ES5/AMD) and 
(b) global object property definitions.
Implied globals are identified
through analysis of AST nodes that 
represent assignment expressions.
Specifically, the variable in the left-hand side
of an assignment is sought in the sets of 
the variables/functions defined
within the scope hierarchy formed 
by the assignment's scope and its surrounding scopes.
The variable comprises an implied global
if it is not declared in the hierarchy. 
Finally, top-level declarations in non-modular ES5
are identified by locating function and variable declaration nodes
which do not have function declaration ancestors.

Let $m_i$ be a module and
$G_i$, $O_i$, $T_i$ refer, respectively
to its global variables,
global object properties and top-level declarations 
identified in its AST.
The resolution of module features from global declarations
involves the following steps:
\begin{enumerate}[label=\emph{Step \arabic*.},leftmargin=1.2cm]
  \item Evaluate global declaration preconditions (see Section~\ref{sect:preconditions}) 
  on $G_i$, $O_i$, $T_i$. In case of any violation return $F_i = \emptyset$.
  \item For each identified global object property $o_k \in O_i$ 
  search for its presence in the feature set of other modules
  \begin{enumerate}
    \item If $o_k$ is also defined in module $m_j$, 
    allocate $o_k$ to the module with lower number of imports,
    so as to normalize coupling among modules,
    \item If not found, then $F_i = F_i \cup \{o_k\}$.
  \end{enumerate}
  \item Return $F_i = F_i \cup G_i \cup T_i$.
\end{enumerate}

In case that global declarations' resolution 
returns $F_i = \emptyset$ for any module $m_i$, 
then the refactoring is abandoned.

\subsubsection{Resolution of fine-grained module dependencies} 
\label{sect:resolve-module-deps}

The resolution of the 
fine-grained module dependencies
involves the identification of the module features that 
each module needs to import from other modules of the same project.
Moreover, it identifies dependencies to external library modules
that are treated differently for compatibility reasons.

Let $d_j\in D$ be a module dependency
from module $m_i$ to a project module $m_t$.
A module dependency represents the import and use
in a module $m_i$ of a module feature $f_{t,k} \in F_t$,
i.e., exported from module $m_t$.
A module dependency is, also, characterized
by the type of use $u_j \in \{R, W, C, L\}$ of $f_{t,k}$ within $m_i$.
If $f_{t,k}$ is only read or called in $m_i$,
then $u_j$ has a value of $R$ or $C$, respectively,
If $f_{t,k}$ is, also, defined in $m_i$
then $u_j = W$.
Thus, a module dependency is defined
as a tuple $d_j = (m_i, m_t, f_{t,k}, u_j)$.
In case that imported module $m_t$ 
corresponds to an external library,
the dependency is represented as
$d_j = (m_i, m_t, \emptyset{}, L)$.

In CommonJS, we identify imported modules
through AST traversal and search for \codew{require} invocations.
A typical \codew{require} invocation for a module $m_j$
is equivalent to an ES6 namespace import,
i.e. the module object exported by $m_j$
is returned and assigned to a variable 
that provides access to the entire feature set $F_j$ of $m_j$.
A specific property of the imported module $m_j$
may also be directly assigned to a variable,
through property access on \codew{require} result.
This coding idiom emulates an ES6 named import in CommonJS
and the receiving variable provides access
to a single feature $f_k \in F_j$.
Finally, a module may be imported only for its side effects,
i.e., the \codew{require} result is not assigned to any variable.

In the case of AMD, 
we identify imported modules 
by analyzing the parameters of the \codew{define}
module declaration function.
Specifically, the first -- optional -- parameter
of the \codew{define} invocation provides a list 
with the relative paths or 
identifiers of imported modules.
Each module $m_j$ is resolved 
and the module object is provided
as parameter to the module definition function
(second parameter of the \codew{define} invocation).
AMD imports are namespace imports,
since parameters of the module definition function
provide access to the imported modules' entire feature set.

The identification of module dependencies for each module $m_i$
is based on the following algorithm:
\begin{enumerate}[label=\emph{Step \arabic*.},leftmargin=1.2cm]
  \item For each imported system module $m_j$ (AMD/CJS)
  identify the subset of its module features $A_j \subseteq F_j$ used in $m_i$
  \begin{enumerate}
    \item Let $u_k$ be the type of use of each $a_k \in A_j$,
    \item For each $a_k \in A_j$ add a module dependency,
    i.e. $D = D \cup \{(m_i, m_j, a_k, u_k)\}$.
  \end{enumerate}
  \item For each imported library module $m_j$ (AMD/CJS)
  add a module dependency, i.e. $D = D \cup \{(m_i, m_j, \emptyset{}, L)\}$.
  \item For each access to a global declaration $g_k \notin F_i$
  \begin{enumerate}
    \changeStart
    \item Let $u_k$ be the type of use of $g_k$.
    Identify the module $m_j$ that encapsulates the global declaration,
    i.e., $g_k \in F_j$,
    \changeEnd
    \item Add a new module dependency $D = D \cup \{(m_i, m_j, g_k, u_k)\}$.
  \end{enumerate}
\end{enumerate}

\subsubsection{MDG construction algorithm} \label{para:MDGcaseStudy}

On the basis of the aforementioned definitions and algorithms
the construction of the MDG data structure for an ES5 project
is summarized below:

\begin{enumerate}[label=\emph{Step \arabic*.},leftmargin=1.2cm]
  \item Initialize sets $M = \emptyset$, $D = \emptyset$
  \item For each source file $f_i$ with name $n_i$
  \begin{enumerate}
    \item Evaluate module format preconditions on $f_i$ (see Section~\ref{sect:preconditions}).
    In case of any violation, refactoring of $f_i$ is abandoned.
    \item Resolve the module structure $F_i$.
    If $F_i \neq \emptyset$ then $M = M \cup \{(n_i, F_i)\}$
    else refactoring of $f_i$ is abandoned.
  \end{enumerate}
  \item For each module $m_i \in M$
  \begin{enumerate}
    \item Resolve module dependencies 
    $D_i$ against modules $M' = M \setminus \{m_i\}$,
    \item $D = D \cup D_i$
  \end{enumerate}
  \item Return $(M, D)$.
\end{enumerate}

\begin{figure}[hbt]
   \centering
   \captionsetup{justification=centering}
   \includegraphics[scale=0.6]{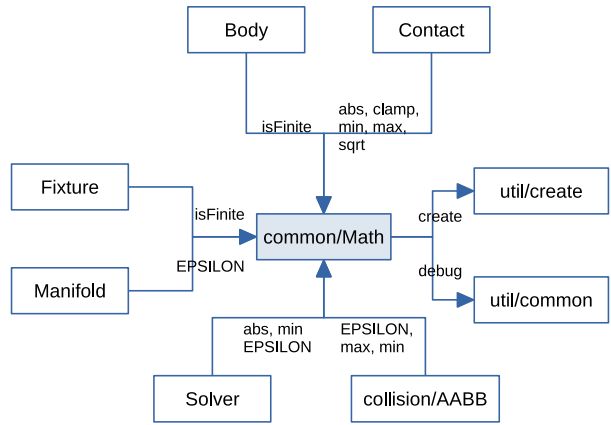}
   \caption{Module Dependence Graph fragment from \planckjs{} project.}
   \label{fig:mdg-example}
\end{figure}

Figure \ref{fig:mdg-example} 
presents a fragment of the MDG
constructed during this stage
for the project \planckjs{}.
The fragment focuses on module \codew{Math},
its outgoing dependencies
and a subset of its incoming dependencies from other modules.
The module's outgoing dependencies 
correspond to the module features that 
are imported and used in the module, 
while incoming dependencies 
correspond to the module's features that 
are exported to client modules. 
For instance, \codew{Math} imports 
the feature \codew{create} from 
the module \codew{create}, 
while it exports \change{the feature \codew{isFinite} 
to the client modules \codew{Fixture} and \codew{Body}}.

\subsubsection{Refactoring preconditions} \label{sect:preconditions}

Refactoring preconditions are evaluated
by the refactoring identification algorithm 
on each module of the analyzed system.
They prevent evaluation errors in the refactored code
and ensure that the external behaviour of the system is preserved.
The identified preconditions,
depending on their scope of application,
are grouped in three categories: 
(a) global declarations' preconditions, 
(b) module object destructuring preconditions,
(c) module format preconditions.

\emph{Global declarations' preconditions} involve 
defining global variables and 
top-level functions with the same name 
in different modules:

\begin{enumerate}
    \item Each global variable should 
    be introduced in a single location. 
    Conflicting declarations 
    across different files are 
    a potential source of errors.
    
    \item In non-modular ES5 or AMD,
    the names of top-level functions 
    should be unique across all JavaScript files.
    Different top-level functions with the same name 
    in multiple modules lead to function overriding;
    the binding in the module's namespace
    corresponds to the most recently loaded function.
\end{enumerate}

\emph{Module object destructuring preconditions}
involve extracting the properties of 
the module object that 
can be independently exported, 
while preserving the program's external behaviour:

\begin{enumerate}

    \item Object properties should be defined
    with dot notation.
    Properties defined with bracket notation 
    may lead to inaccurate conclusions 
    regarding their declaration module,
    in cases that the 
    object is indexed by a variable.
    
    \item Module objects should not 
    be fully referenced  
    (e.g. provided in function invocations). 
    Detaching object properties from 
    these objects may lead to errors 
    originating from name aliasing.
    
    \item Module objects should not be modified. 
    Detaching properties from 
    such objects may lead to errors 
    related to violations of the ES6 specification~\cite{Rauschmayer15}.
    
\end{enumerate}

\emph{Module format preconditions} involve 
restrictions that are imposed by 
the language specification. 
They, also, involve restrictions 
that are not imposed by ES5 formats, 
but are introduced in the ES6 specification. 
The violation of these restrictions may 
introduce defects in the refactored code or 
affect the program's external behaviour:

\begin{enumerate}

    \item Function definitions that reference \codew{this}
    should be declared in files that 
    enforce strict mode.
    In case that a function is 
    neither an object method nor a constructor,
    the value of \codew{this} is determined by 
    the module's code mode. 
    In \emph{non-strict} mode, 
    \this{} references the global object.
    In \emph{strict} mode 
    \this{} is \codew{undefined},
    as its value is not updated during 
    the function invocation \cite{mdnJSRef}.
    Since ES6 modules enforce 
    \emph{strict} mode \cite{Rauschmayer15},
    the transformation of an ES5 source file in 
    \emph{non-strict} mode into an ES6 module
    would lead to runtime errors 
    originating from the value of \codew{this}.
    
    \item Invocations of \codew{require} in CJS 
    modules should be located in the module's top-level scope.
    Notice that in ES6,
    the introduction of static imports and exports 
    is allowed only at the module scope.
    In-place transformation of 
    \codew{require} calls to ES6 static imports 
    would lead to evaluation errors, 
    if they are nested in block statements \cite{Rauschmayer15}. 
    This, also, applies to the replacement of return statements 
    with export statements in AMD.
    
\end{enumerate}


\subsection{Source Code Transformation Procedure} 
\label{sourceCodeTransformationProcedure}

The source code transformation
for migration to ES6 modules
employs the MDG constructed in 
the refactoring candidate identification stage.
Each MDG node, corresponding to a module $m_i$, 
is individually refactored in the following steps:

\paragraph{Step 1}If $m_i$ is an AMD module, clear its dependencies to the AMD API:
  \begin{enumerate}[label=(\alph*)]
    \item Replace the invocation of the \codew{define} function
    with a list of all statements 
    included in the body of the module definition function.
    \item Replace the return statement 
    included in the statement list
    with a variable assignment of the form: 
    \code{var mod_<filename> = <return_expr>},
    where \code{<return_expr>} is the expression
    returned by the replaced statement.
   \end{enumerate}

\begin{figure*}[hbt!]
 \centering
 \captionsetup{justification=centering}
 \includegraphics[scale=0.7]{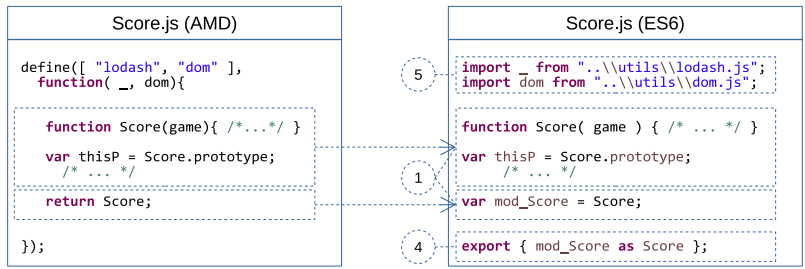}
 \caption{Module definition and exported feature migration in AMD modules.}
 \label{fig:module-migration-amd}
\end{figure*}

\begin{figure*}[hbt!]
 \centering
 \captionsetup{justification=centering}
 \includegraphics[scale=0.7]{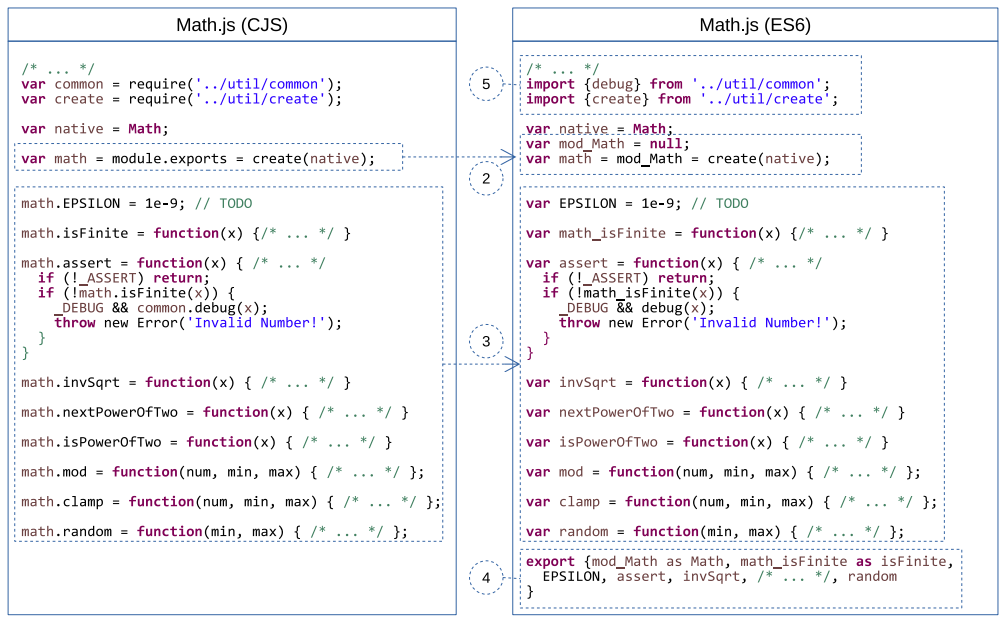}
 \caption{Migration of module exports syntax in CJS modules.}
 \label{fig:mathJsRef}
\end{figure*}

The application of this step to an AMD module
is illustrated in Figure~\ref{fig:module-migration-amd}.
The affected statements are enclosed in a dashed frame
and are marked with the step number (1).
In the rest of this section,
numbers inside parentheses will be used
to refer to similarly marked parts 
of Figures~\ref{fig:module-migration-amd},~\ref{fig:mathJsRef}.

\paragraph{Step 2}If $m_i$ is a CJS module, 
	clear dependencies to the CommonJS API:
  \begin{enumerate}[label=(\alph*)]
    \item Declare an empty placeholder variable for the module object
    with the following naming scheme:
    \code{var mod_<filename> = null}.
    \item Replace all references to \codew{module.exports} 
    \change{or its alias \codew{exports}} with the above variable.
    \item Eliminate all \codew{require} statements.
  \end{enumerate}

Figure~\ref{fig:mathJsRef} presents
the application of this step
to the \codew{Math.js} module of the \codew{planck.js} project.
The newly introduced variable \codew{mod\_Math},
replaces \codew{module.exports}
and holds the reference to the module object
in refactored code (2). 
Variable \codew{math} remains an alias to the module object.

\paragraph{Step 3}Apply module object destructuring 
  by processing each identified feature $f_j \in F_i$  
  for the module $m_i$.
  Each feature $f_j$ is declared and initialized in ES5
  through a property-value pair in object literal notation
  or through an assignment statement
  that binds a property to the module object.
  In order to extract $f_j$ from the module object:
  \begin{enumerate}[label=(\alph*)]
    \item Introduce a variable declaration for $f_j$
    that has the form:
\begin{lstlisting}[captionpos=l, % invalid input hides the caption !!!
      numbers = none]
var <feature> = <feature_expr>
\end{lstlisting} 
    where \code{<feature>} corresponds to the name of $f_j$.
    In case of name conflicts,
    the feature name is prefixed
    with the module name, i.e., \code{<file_name>_}.
    The term \code{<feature_expr>} corresponds
    to the expression used for initialization of $f_j$
    during its declaration.
    \item Replace references to $f_j$ in module $m_i$
     with the feature name.
    \item Remove the property-value pair
    or the property binding assignment statement,
    originally used for $f_j$ declaration.
  \end{enumerate}

The declaration of module features 
in the ES5 version of \codew{Math.js}
(left part of Figure~\ref{fig:mathJsRef}),
is based on property binding assignments 
to the \codew{math} alias of the module object.
All property bindings can be extracted
as accordingly named variable declarations  (3),
with the exception of \codew{isFinite}
that introduces name collision
with a global JavaScript function.

\paragraph{Step 4}Declare the exported module features of $m_i$
through evaluation of all its incoming MDG dependencies.
  \begin{enumerate}[label=(\alph*)]
    \item Initialize exported features $E_i = \emptyset$
    \item For each incoming MDG edge $(m_k, m_i, f_j, u_j)$,
    originating from a client module $m_k$,
    add $f_j$ to the set of exported features $E_i$.
    In case that the client module modifies the value of $f_j$,
    i.e., $u_j = W$,
    declare a mutator function for $f_j$
    and add it to exported features $E_i$.
    The following code listing provides a template
    for mutator function declaration.

\begin{lstlisting}[captionpos=l, % invalid input hides the caption !!!
      numbers = none]      
function set_<feature>(value) { <feature> = value; }
\end{lstlisting}
    \item Declare an export statement listing all exported features in $E_i$.
    For name consistency, provide an alias and restore the original name
    of all features that were prefixed for handling name conflicts.
  \end{enumerate}

The AMD module of Figure~\ref{fig:module-migration-amd}
provides a single export (4),
since the module object could not be further destructured.
On the other hand, 
the CJS module of Figure~\ref{fig:mathJsRef},
exports a list of features in a single statement (4).
In both cases, renamed features are restored to their original version. 

\changeStart
Notice that Step 4 may be adapted
in case that the analyzed project is not a standalone application,
e.g., it implements a reusable library.
The reason is that the part of a library's public API
without incoming MDG dependencies in the codebase,
will be accidentally encapsulated
and become unavailable to its clients.
Step 4, in this case, exports all features
$f_j \in F_i$ for each module $m_i$. 
However, if the public API is invoked in test code,
Step 4 can be applied without modifications,
as test code may provide the required incoming dependencies. 
\changeEnd

\paragraph{Step 5}Establish module dependencies for $m_i$
through evaluation of all its outgoing MDG dependencies.
\begin{enumerate}[label=(\alph*)]
    \item For each outgoing MDG edge $(m_i, m_k, f_j, u_j)$
    to a module feature $f_j$ of a system module $m_k$,
    introduce an appropriate named import.
    The name of $f_j$ (\code{<feature>}) is maintained,
    or in case of name conflicts
    it is prefixed with a module identifier (\code{<prefix>}).
    The identifier corresponds to the file name of $m_k$,
    or if the conflict persists it is further prefixed
    with the name of the parent folder of the module declaration and so on.
    In case that the imported feature is defined in $m_i$, i.e., $u_j = W$,
    the corresponding mutator function is, also, imported. 
    The introduced named imports follow the pattern:
\begin{lstlisting}[captionpos=l, % invalid input hides the caption !!!
      numbers = none]      
import {<feature> [as <prefix>_<feature>]} from <file>;
\end{lstlisting}
    \item For each imported feature $f_j$,
    replace all references to it with the feature name.
    In case of assignment statements
    that define the value of $f_j$,
    they are replaced with an invocation
    of the feature's mutator function.
      
    \item For each outgoing MDG edge $(m_i, m_k, \emptyset{}, L)$
    to an external library module, 
    introduce a default import, i.e.,
\begin{lstlisting}[captionpos=l, % invalid input hides the caption !!!
      numbers = none]      
import <lib> from <file>;
\end{lstlisting}
	where \code{<lib>} corresponds to the variable name
	used to reference the library module in ES5 code. 
\end{enumerate}

Figure~\ref{fig:module-migration-amd}
presents the import of two library modules
through \emph{default} import statements (5).
In Figure~\ref{fig:mathJsRef},
the \codew{require} statements
are replaced by two named imports
that introduce the minimum required module features
to the namespace of \codew{Math.js} (5).

\section{Empirical evaluation} 
\label{evaluation}

We have designed and performed
an experimental study 
on the proposed method for automated
refactoring of ES5 software projects 
towards ES6 fine-grained modularity.
The \emph{goal} of this study is 
to analyze the proposed method
for the \emph{purpose} of evaluation
with respect to effectiveness and practicality
from the \emph{perspective} of
a developer, that considers the implications of the refactoring
and its potential for automation,
in the \emph{context} of a set of open source projects.

\subsection{Research Questions}

The experimental study aims at answering
the following research questions:

\changeStart
\emph{RQ1. How prevalent is the presence of ES5 code, modular or not, in real-world applications?}
At first, we investigate to which extent ES5 code,
that is either non modular or supports the AMD/CJS module formats,
is still prevalent in web applications.
The results of this study will provide empirical evidence
on the scope and practicality of the proposed method.
\changeEnd

\emph{RQ2. Is there a developer intent 
for independently reusable features within modules?}
Our purpose is to explore at which extent
the proposed refactoring to ES6 fine-grained modularity
is aligned with the semantics of the analyzed modules
and the developer's intent 
for independently reusable features within modules.
The results will support the practicality
of the proposed refactoring for effective usage of ES6 language
constructs after migration of legacy code.

\changeStart
\emph{RQ3. What are the effects of the proposed refactoring to system modularity?}
Our aim is to study the level of granularity, 
in terms of individually exported module features,
that can be introduced in project modules
and the potential for encapsulation of module features
within their declaring modules.
The research question, also, evaluates potential reduction to module coupling
due to fine-grained dependencies introduced by ES6 named imports.
The results of this research question will highlight
the effectiveness of the proposed method
for increasing the granularity of reusable module features
and reducing the internal coupling of system modules. 
\changeEnd

\emph{RQ4. Does the proposed source code transformation
preserve the external behaviour of the analyzed system?}
The research question seeks to empirically investigate
the soundness of the proposed refactoring,
i.e., whether the application of the refactoring
does not introduce syntax errors or
alter the external behaviour of refactored code.
The results of this study will further support
the practicality of the proposed method.

\subsection{Context Selection}

The context of the experimental study comprises 19 JavaScript projects. 
The selection of these projects is based on 
the following criteria:
(a) they are open source, for study replication reasons,
(b) they are implemented in ES5 or below,
(c) they are either non-modular or use the AMD/CJS module formats 
\change{as implemented in RequireJS/Node.js respectively},
(d) their test suite, if available, can be successfully executed,
\changeStart
(e) to the extent possible, they are used as benchmarks in earlier work in this area,
 e.g.,~\cite{Vazquez2018}.
\changeEnd

\begingroup
\setlength{\tabcolsep}{6pt} 
\renewcommand{\arraystretch}{1.2} 
\begin{table*}[htb]
\centering
 \scriptsize
\begin{tabular}{l l l l l l}
\toprule
\multicolumn{1}{c}{\textbf{Project}} &
\multicolumn{1}{p{1.5cm}}{\raggedright\textbf{Module Format}} &
\multicolumn{1}{c}{\textbf{SLOC}} &
\multicolumn{1}{p{2cm}}{{\color{black}\raggedright\textbf{Test Coverage}}} &
\multicolumn{1}{c}{\textbf{Tag/Commit}} &
\multicolumn{1}{c}{\textbf{Github Repository}}\\ \hline

UltraTetris & \multirow{4}{1.5cm}{--} & 326  & {\color{black}-} & master@5ad237e & {silviolucenajunior/UltraTetris} \\
Hangman     &  & 346 & {\color{black}-} & master@b950842 & {aurobindodebnath/Hangman} \\
TicTacToe   &  & 919 & {\color{black}-} & master@346ebe8 & {seanpr/TicTacToe} \\
uki         &  & 2360   & {\color{black}-} & master@6cd2e47 & {crcx/uki} \\
\hline
GluttonousSnake  & \multirow{5}{1.5cm}{AMD}   & 378 & {\color{black}-} & master@c6b49cb & {yyfer/GluttonousSnake}  \\
astix            &  & 481 & {\color{black}-} & master@f0ecc39 & {migace/astix} \\
game-of-life     &  & 887 & {\color{black}-} & master@1d83874 & {devlysh/game-of-life} \\
tetrisJS         &  & 1370 & {\color{black}-} & master@27712a0 & {marneborn/tetrisJS} \\
dynablaster-js-port & & 3679 & {\color{black}-} & master@5b5052e & {gardziej/dynablaster-js-port} \\
\hline
backbone-tableview  & \multirow{10}{1.5cm}{CommonJS} & 279 & {\color{black}\pct{100.0}} & master@0c26357 & {mbrevda/backbone-tableview} \\
easystarjs          &   & 369 & {\color{black}\pct{100.0}}   & v0.4.3 & {prettymuchbryce/easystarjs} \\
geojsonhint         &   & 473 & {\color{black}\pct{100.0}}    & v.2.0.0 & {mapbox/geojsonhint} \\
express-session     &   & 612 & {\color{black}\pct{99.0}}   & v1.15.6 & {expressjs/session}  \\
underscore.string   &   & 873 & {\color{black}\pct{99.3}}  & 3.3.4 & {esamattis/underscore.string}    \\
messy               &   & 1694 & {\color{black}\pct{91.8}}   & 6.11.0 & {papandreou/messy} \\
virtual-dom         &   & 1978 & {\color{black}\pct{97.5}}   & v2.1.1 & {Matt-Esch/virtual-dom} \\
recipe-parser       &   & 4134 & {\color{black}\pct{41.0}}  & master@626f124 & {catesandrew/recipe-parser} \\
planck.js           &   & 10390 & {\color{black}\pct{40.5}}  & v0.2.7 & {shakiba/planck.js} \\
goojs               &   & 57373 & {\color{black}\pct{48.0}}  & v0.16.8 & {GooTechnologies/goojs} \\

\bottomrule
\end{tabular}

\captionsetup{justification=centering}
\caption{JavaScript projects used in the experimental study.}
\label{Tab:projectImplementationDetails}
\end{table*}
\endgroup

Table \ref{Tab:projectImplementationDetails} presents 
the projects used in our evaluation, 
along with relevant implementation details.
Column 2 provides the module format used in each project.
Notice that the first four projects are non-modular.
Column 3 (SLOC) includes the total lines of JavaScript production code, 
computed with CLOC\footnote{https://github.com/AlDanial/cloc}.
In case that the codebase of a project includes external libraries,
library code is excluded from further analysis.
The reason is that these libraries
make use of dynamic features of the language
(e.g. on-the-fly creation and deletion of object properties, 
dynamic code generation).
These features undermine the results of 
static analysis \cite{Feldthaus13} and, thus,
their refactoring may introduce errors or
influence the program's external behaviour.
\changeStart 
Column 4 provides the test coverage of each project, 
in terms covered statements by its test suite. 
Test coverage is not reported
for projects without a test suite.
\changeEnd
Column 5 specifies the tag/commit 
of the project's revision history
that was used in this study
and Column 6 provides the projects'
Github repository names. 

\changeStart
The proposed method is implemented 
as an open source Node.js tool
\footnote{https://github.com/katerinapal/es52es6}, 
which employs \jscodeshift\footnotemark{}
for Abstract Syntax Tree (AST) traversal and transformation, 
JSHint\footnotemark[\value{footnote}]
for the resolution of implied globals
and Tern\footnotemark[\value{footnote}]
for the dataflow analysis 
required in the module dependency identification step 
(\ref{sect:resolve-module-deps}). 
The JavaScript code that 
is embedded in the HTML files of the analyzed web applications
is extracted with the help of 
the jsoup HTML parser\footnotemark[\value{footnote}]
\changeEnd 
The implemented prototype requires 
limited user interaction, 
since the user only needs to specify 
the application to be refactored 
and the format that is used for its implementation. 
The evaluation of the prototype
is performed on a workstation, 
equipped with an Intel-Core i7-7700 CPU @3.60 Ghz 
and 16GB RAM.

\footnotetext{jscodeshift: https://github.com/facebook/jscodeshift, JSHint: https://github.com/jshint /jshint, 
Tern: https://github.com/ternjs/tern, jsoup: https://jsoup.org}

\subsection{Evaluation Results}

\changeStart
\subsubsection{RQ1. How prevalent is the presence of ES5 code, 
modular or not, in real-world applications?}

The first research question explores
how relevant is the proposed method 
with the needs of current web applications.
Since the ES6 (ES2015) specification
has been fully implemented in most JavaScript runtimes,
a question naturally arises as to what extent
web applications still use the legacy module formats
and have not yet migrated to ES6.
To the best of our knowledge, 
there is no published study that explores
the use of AMD/CJS formats and 
the current adoption of ES6 modules. 

In order to answer RQ1,
we analyzed the code-base of all Github repositories
that use JavaScript as their main language
and have received at least 
a minimum of developer engagement in the recent years. 
We acquired access to the source code of the main branch 
and the public activity timeline
of Github repositories,
through the \emph{Github Activity Data}
\footnote{https://console.cloud.google.com/marketplace/product/github/github-repos}
and \emph{GithubArchive} 
\footnote{https://www.gharchive.org/}
public BigQuery datasets. 
Among approximately 4 million repositories that include JavaScript code,
we narrowed down our analysis to 14,167 repositories 
on the basis of the following criteria: 

\begin{itemize}
  \item the primary language of the repository, in terms of bytes of code, is JavaScript or HTML. 
  We excluded repositories that include Java, Objective-C or C\# code,
  in order to avoid analyzing cross-platform mobile applications, 
  where migration to ES6 is constrained by application development frameworks. 
  
  \item each repository has received at least 10 commits and 5 stars since 01.01.2018.
  Our purpose is to exclude abandoned projects 
  that would bias our results in favor of legacy module formats. 
     
\end{itemize} 

In a next stage,
we collected all JavaScript files from the aforementioned repositories
and classified them as AMD, CJS or ES6 modules.
The classification is based on regular expressions
that match the syntax of each module format
in the source code of each file.
Regular expressions are appropriately formulated
so that each file is categorized to a single module format
and are executed against a BigQuery table with JS file contents.
We excluded from our analysis build tool configuration files
(e.g. webpack, rollup),
as well as source files located at the installation folder
of package dependencies (e.g., node\_modules).

The results of JS content classification to module formats
are available in Table~\ref{Tab:jsmodules-distr}.
The table provides the number of JS files (Column 2),
among all 14,167 repositories,
that declare AMD, CommonJS, ES6 modules
or do not use any standard module format.
Column 3 estimates the share of all files
that fall under these categories.
Classification results show that
a large share of JS content (\pct{31.3})
is either non-modular or does not use a standard module format.
Moreover, about one-third of all JS files (\pct{35.4})
use the standard ES5 module formats,
AMD (\pct{6.7}) and CJS (\pct{28.7}).
Finally, the adoption of the ES6 module format
characterizes another one-third of all files (\pct{33.3}).

\begingroup
\setlength{\tabcolsep}{6pt} 
\renewcommand{\arraystretch}{1.2} 
\begin{table*}[htb]
\centering
 \scriptsize
\begin{tabular}{l r r}
\toprule

\textbf{Module format} 
& \multicolumn{1}{c}{\textbf{Files}}
& \multicolumn{1}{c}{\textbf{\%}}\\ 
\hline
AMD			& 65,457 	& 6.7	\\
CommonJS 	& 281,874 	& 28.7 	\\
ES6			& 327,367	& 33.3 	\\
Other 		& 307,523	& 31.3 	\\
\hline
Total		& 982,221	& 100.0 \\

\bottomrule
\end{tabular}

\captionsetup{justification=centering}
\caption{Classification of JS files to module formats.}
\label{Tab:jsmodules-distr}
\end{table*}
\endgroup

\begin{answer}
Legacy module formats and non-modular ES5 code 
are still widely used in actively maintained open source projects.
This stresses the need for methods and tools
for their automated migration to ES6 modules
in order to reap the benefits of new language features.
\end{answer}
\changeEnd

\subsubsection{RQ2. Is there a developer intent 
for independently reusable features within modules?}

The second research question is related to
the practicality of the proposed refactoring.
It's purpose is to investigate whether
the extracted fine-grained module features
correspond to elements with similar semantics
in the original module design, i.e.,
elements that belong to the module scope and 
may be reused independently from each other.
Thus, the language constructs (ES6 named imports/exports)
that are automatically applied through refactoring, 
could have been employed by the programmer 
if they were supported by ES5.
  
In order to answer RQ2,
we conducted an exploratory study 
on frequently occurring patterns 
on the design of module objects.
The context of this study comprised 
the subset of benchmark projects
that use the AMD/CJS formats.
Non-modular ES5 projects were excluded,
since their source files do not declare module objects.
The study involved, at first, 
the identification of common patterns 
through code inspection on a sample of modules
from each benchmark project.
Then, we estimated the frequency of each pattern,
across all modules,
through a static analysis tool
implemented for that purpose.
Besides the \emph{namespace object} pattern,
already described in~\ref{sect:resolve-mod-features},
code inspection revealed several module object patterns
that can be grouped in two categories:
(a) \emph{factory} objects and (b) \emph{utility} objects.
The categorization focuses on the purpose of module objects
and abstracts variations in their design.

A \emph{factory object} is a function that
implements a template for object creation and initialization.
The function is usually a \emph{function constructor}, 
a factory method or a generic object configuration function.
Its purpose is to initialize a new object 
with attributes and methods
through statements in the function body
(property bindings and property assignments 
on \this{} reference).
Instance methods are, also, attached
through property bindings
to the factory object's \codew{prototype} object.
Moreover, the module code, usually, 
includes statements that apply property bindings
directly on the factory object.
These bindings are extracted as independent module features
by the proposed refactoring.
They represent static members of the class declaration
that is emulated within the module,
and can be reused independently from the instances
created by the factory object.

Figure~\ref{fig:factory-object} presents a factory object
declared in \codew{Vec2} module of \planckjs{} project.
The module object acts as a function constructor
or as a factory method, 
depending on being invoked 
with the \codew{new} operator or not.
An instance method of \codew{Vec2} objects
is declared in lines 21--23.
On the other hand, direct property bindings to \codew{Vec2} object,
e.g., \codew{zero}, \codew{neo},
represent static utility methods that can be extracted
and exported as individual module features.

\begin{figure}[htb]
  \begin{lstlisting}[language=JavaScript,
      basicstyle=\ttfamily\scriptsize,
      captionpos=l, % invalid input hides the caption !!!
      xleftmargin=2em, numbers = left]
/* ... */
module.exports = Vec2;
/* ... */
function Vec2(x, y) {
  if (!(this instanceof Vec2)) {
    return new Vec2(x, y);
  }
  if (typeof x === 'undefined') {
    this.x = 0;
    this.y = 0;
  }
  /* ... */
}

Vec2.zero = function() {/*...*/};

Vec2.neo = function(x, y) {/*...*/};

Vec2.clone = function(v) {/*...*/};

Vec2.prototype.toString = function() {
  return JSON.stringify(this);
};

Vec2.isValid = function(v) {/*...*/}

Vec2.assert = function(o) {/*...*/};
/* ... */
\end{lstlisting}
\caption{Factory object in \codew{Vec2} module of \planckjs{} project.}
\label{fig:factory-object}
\end{figure}

In the case of \emph{utility objects},
the module object is a non-empty function 
that serves a domain-specific functionality
and does not include bindings to the \this{} reference.
Furthermore, its prototype object is not extended
within the module scope.
Bound properties to this module object are, also, extracted
and exported as separate module features after refactoring.
However, due to the generic structure of \emph{utility objects},
we cannot generalize on the role of extracted module features
in module design.
Thus, we assume 
that the individual reuse of these module features
might not express the original intent of the developer.  

\begingroup
\setlength{\tabcolsep}{6pt} 
\renewcommand{\arraystretch}{1.2} 
\begin{table}[htb]
\centering
\scriptsize
\begin{tabular}{@{} l r r r @{}}
\toprule
\multicolumn{1}{p{0.5cm}}{\textbf{Project}} &
\multicolumn{3}{c}{\textbf{Module Objects}} \\
\cmidrule{2-4} &
 \multicolumn{1}{c}{Factory} & 
 \multicolumn{1}{c}{Namespace} & 
 \multicolumn{1}{c}{Utility}\\ \hline

GluttonousSnake     & \valpct{1}{25.0} & \valpct{3}{75.0} & \valpct{0}{ 0.0} \\ 
astix               & \valpct{0}{ 0.0} & \valpct{8}{ 100} &  \valpct{0}{ 0.0} \\ 
game-of-life        & \valpct{4}{66.6} & \valpct{1}{16.7} & \valpct{1}{16.7} \\ 
tetrisJS            & \valpct{5}{50.0} & \valpct{5}{50.0} & \valpct{0}{ 0.0} \\ 
dynablaster-js-port & \valpct{38}{80.9} & \valpct{5}{10.6} & \valpct{4}{ 8.5} \\ \hline 
backbone-tableview  & \valpct{1}{ 8.3}  & \valpct{1}{ 8.3} & \valpct{10}{83.4} \\ 
easystarjs          & \valpct{2}{66.7}  & \valpct{1}{33.3} & \valpct{0}{ 0.0} \\ 
geojsonhint         & \valpct{0}{ 0.0}  & \valpct{0}{ 0.0} & \valpct{3}{ 100} \\ 
express-session     & \valpct{4}{80.0}  & \valpct{1}{20.0} & \valpct{0}{ 0.0} \\ 
underscore.string   & {\color{black}\valpct{1}{ 1.0}}  & {\color{black}\valpct{5}{ 5.2}} & {\color{black}\valpct{90}{93.8}} \\ 
messy               & \valpct{10}{62.5} & \valpct{2}{12.5} & \valpct{4}{25.0} \\ 
virtual-dom         & \valpct{7}{22.6}  & \valpct{1}{ 3.2} & \valpct{23}{74.2} \\ 
recipe-parser       & \valpct{1}{ 6.7}  & \valpct{2}{13.3} & \valpct{12}{80.0}   \\ 
planck.js           & {\color{black}\valpct{37}{74.0}} & {\color{black}\valpct{4}{8.0}} & {\color{black}\valpct{9}{18.0}}    \\ 
goojs               & \valpct{433}{81.8} & \valpct{74}{14.0} & \valpct{22}{4.2}  \\ \hline 
\textbf{Total}      & {\color{black}\valpct{544}{65.1}} & {\color{black}\valpct{113}{13.5}} & {\color{black}\valpct{178}{21.3}} \\ 
\bottomrule
\end{tabular}

\captionsetup{justification=centering}
\caption{Common patterns of module objects and their frequency.}
\label{Tab:moduleStructure}
\end{table}
\endgroup

Table~\ref{Tab:moduleStructure} presents static analysis results
on the classification of all module objects in AMD/CJS projects
to module pattern categories.
Specifically, Columns 2--4 provide for each project
the number of module objects that fall, respectively,
into the factory, namespace and utility object categories. 
The results show that the majority of module objects, 
i.e., \change{\pct{65.1}}, are classified as factory objects.
Moreover, a smaller but significant share of module objects (\pct{13.5})
are classified as namespace objects,
while utility objects correspond to \change{\pct{21.3}} of 
all module objects.
Thus, a total of \change{\pct{78.6}} of module object declarations
follow the \emph{namespace} or \emph{factory} object pattern,
i.e., their module scope emulates,
either a container for features or a class.
In each one of these declarations,
the module object's bound properties
represent features that reside semantically at the module scope
and can be individually reused as ES6 named exports. 
Bound properties in the rest of module declarations, 
i.e, in \emph{utility} objects,
may also be reusable as individual features,
but their relevance must be inspected 
on a case-by-case basis. 

\begin{answer}

Bound properties in \change{\pct{78.6}} of module object declarations
correspond to namespace elements or
static features of an emulated class,
revealing a developer intent
for their individual reuse at module level.
\end{answer}

\subsubsection{RQ3: 
What are the effects of the proposed refactoring to system modularity?}

In order to answer RQ3, 
we evaluate the module structure after refactoring,
where module objects are destructured and replaced 
by individually reusable module features.
We investigate the reusability of module features across modules,
\change{as well as the impact of fine-grained dependencies to module coupling}.

\begingroup
\setlength{\tabcolsep}{2pt} 
\renewcommand{\arraystretch}{1.2} 
\begin{table}[htb]
\centering
\scriptsize
\begin{tabular}{@{\extracolsep{1pt}}l r r r r}
\toprule
\multirow{2}{0.8cm}{\textbf{Project}} &
\multicolumn{1}{c}{\textbf{Modules}} &
\multicolumn{3}{c}{\textbf{Module Features}} \\
\cmidrule{3-5}
 & & \multicolumn{1}{c}{Total} 
  & \multicolumn{1}{c}{Exported} 
  & \multicolumn{1}{c}{Encapsulated} \\ \hline

UltraTetris         & 5   & 15  & \valpct{12}{80.0}	  & \valpct{3}{20.0} \\
Hangman             & 9   & 44  & \valpct{27}{61.4}	  & \valpct{17}{38.6} \\
TicTacToe           & 2   & 12  & \valpct{5}{41.7}	  & \valpct{7}{58.3}	\\
uki                 & 3   & 39  & \valpct{5}{12.8}	  & \valpct{34}{87.2}	\\ 
\hline
GluttonousSnake     & 4   & 10  & \valpct{8}{80.0}	  & \valpct{2}{20.0}	\\
astix               & 10  & 49  & \valpct{42}{85.7}	  & \valpct{7}{14.3}	\\
game-of-life        & 7   & 23  & \valpct{16}{69.6}   & \valpct{7}{30.4} \\
tetrisJS            & 11  & 41  & \valpct{20}{48.8}   & \valpct{21}{51.2} \\
dynablaster-js-port & 49  & 259	& \valpct{61}{23.6}   & \valpct{198}{76.4}	 \\
\hline
backbone-tableview  & 12  & 19  & \valpct{19}{ 100}	  &  \valpct{0}{ 0.0}	 \\
easystarjs          & 3   & 11  & \valpct{11}{ 100}	  &  \valpct{0}{ 0.0}	 \\
geojsonhint         & 3   & 3   & \valpct{3}{ 100}	  &  \valpct{0}{ 0.0}	 \\
express-session     & 5   & 9   & \valpct{9}{ 100}	  &  \valpct{0}{ 0.0}	 \\
underscore.string   & 97  & 191 & \valpct{166}{86.9}  & \valpct{25}{13.1}	 \\
messy               & 16  & 111 & \valpct{111}{ 100}  & \valpct{0}{ 0.0}	 \\
virtual-dom         & 31  & 45  & \valpct{44}{97.8}	  & \valpct{1}{ 2.2}	 \\
recipe-parser       & 16  & 65  & \valpct{53}{81.5}	  &  \valpct{12}{18.5}	 \\
planck.js           & 55  & 285	& \valpct{276}{96.8}  & \valpct{9}{ 3.2}	 \\
goojs               & 540 & 2045& \valpct{1734}{83.8} & \valpct{332}{16.2} \\
\hline
\textbf{Total}      & 878 & 3276 & \valpct{2601}{79.4} & \valpct{675}{20.6} \\
\bottomrule
\end{tabular}

\captionsetup{justification=centering}
\caption{Exported and encapsulated module features after the refactoring.}
\label{Tab:exportedFeatures}
\end{table}
\endgroup

The effect of the proposed refactoring 
on the granularity of independently reusable elements within a project,
can be inferred by Columns 2, 3 of Table~\ref{Tab:exportedFeatures}.
\changeStart
Column 2 (Modules) provides the number of module objects
declared within each project,
while Column 3 (Total Module Features) sums up all module features, i.e.,
(a) bound properties that are extracted from each module object and
(b) global declarations that are allocated
and accessed through modules after refactoring.
Initially, the units of reuse were
coarse-grained module objects,
each one declared in its own source file.
\changeEnd
The proposed method sets the unit of reuse
at the granularity of module features,
with multiple feature declarations and exports per source file.
The figures in Columns 2 and 3
show that this shift in granularity results
to an increase in the number of reusable elements per project
by a factor of 4 on average.
This increase in fine-grained reusable elements reduces module coupling, 
as will be demonstrated hereafter, 
while supporting application bundling tools 
in effective tree-shaking, i.e., identification and elimination of
unused code from deployment artifacts~\cite{rollupjs,webpackTreeShaking}.

The module features that are, actually, reusable within each project
are provided in Column 4 \change{(Exported Module Features)}.
These module features are required and used
by at least one module, besides their declaring module.
After refactoring,
they are declared as \emph{named exports}
and are reused in client modules
through corresponding \emph{named import} statements.
Moreover, the column provides, inside parentheses,
the respective fraction of reusable module features
against total module features of the project (Column 2).
According to figures in Column 4,
\pct{79.4} of all module features are reusable and should be exported,
while, in 13 out of 19 projects,
these module features exceed \pct{80}.
 
However, a significant part of module features,
especially in non-CommonJS projects,
may not be exported but encapsulated in their declaring modules.
\changeStart
Notice that CJS projects implement Node.js library packages
and, thus, their encapsulated features are estimated
by integrating in the static analysis their unit tests,
in order to act as client code of their public API.
In case that unit tests
do not provide sufficient coverage of the public API,
encapsulation of module features can be disabled.
\changeEnd
As revealed by results in Column 5,
20.6\% of all module features and
14.3\%--87.2\% in non-modular ES5 and AMD projects
are not reused outside their declaring module.
In any case, the results provide evidence 
on the poor support for module encapsulation in ES5.
The proposed refactoring 
omits ES6 named export statements for these features,
restricting their scope and
narrowing the public API of modules.
Thus, it improves modularity 
through better encapsulation.

\changeStart
As concerning the effect of refactoring on module coupling,
we investigate it by comparing the average module coupling of each project
before and after the application of refactoring.
Module coupling estimation is based 
\changeEnd
on the incoming dependencies (\emph{Fan-in}) and
the outgoing dependencies (\emph{Fan-out}) of a module.
Moreover, we estimate \emph{Module Instability}
in analogy to the well-established \emph{Instability} metric
for components or packages~\cite{Martin:2003}.
We draw an analogy between an ES6 module and 
a Java package
\footnote{Or an equivalent construct of another programming
language, e.g. \texttt{C\#} namespace.}
in order to estimate module \emph{Fan-in} and \emph{Fan-out}.
A Java package groups classes (and other types or packages)
and makes them accessible to other packages 
through the \codew{public} class modifier.
Its outgoing dependencies comprise classes of other packages
that are imported by its own classes,
while incoming dependencies comprise classes of other packages
that import one or more of its own classes.
In a similar vein,
an ES6 module groups module features,
that are declared in its source file,
and become accessible to other modules
through export statements.
An ES6 module has outgoing dependencies to module features
that are imported individually or in bulk
(named or default imports)
from other modules.
Incoming dependencies of a module 
comprise the module features of other modules
that import one or more of its own features.

Let $m_i$ be a module, 
$F_i$ its module features
and $D_i$ its set of outgoing dependencies in the MDG,
according to the notation introduced in Section~\ref{refactoringCandidateIdentificationProcedure}.
Moreover, let $F_{i, out}$ be the set of module features
that $m_i$ depends on and imports into its source file
(external library modules are excluded).
The following equation provides a formal specification for $F_{i, out}$:

\begin{fleqn}
\begin{equation*}
\begin{split}
F_{i, out} = 	& \bigcup \{f_{t,k}\ |\ \exists d\in D_i\ s.t. & d=(m_i,m_t,f_{t,k},\_)\ \wedge\ \\
			& & m_t\in M \wedge f_{t,k}\in F_t\} \\
\end{split}
\end{equation*}
\end{fleqn}

The set $F_{i, in}$ of module features that depend on module $m_i$
contains all module features whose declaration
references at least one module feature from $m_i$.
Let $m_s$ be a module with outgoing dependencies
to $m_i$ in the MDG,
i.e., it imports one or more features from $m_i$. 
Moreover, let $A_{i,s} \subseteq F_s$ 
be the module features of $m_s$
that access at least one imported feature from module $m_i$.
We compute $F_{i, in}$ as the union of all sets $A_{i,s}$,
where $m_s$ is a module that imports features from $m_i$.
A formal specification of $F_{i, in}$ is provided below:

\begin{fleqn}
\begin{equation*}
\begin{split}
F_{i, in} = 	& \bigcup A_{i,s}, \text{for each}\ m_s \in M\ s.t. \\
			& (\exists d\in D_s\ |\ d=(m_s,m_i,f_{i,k},\_)\  \wedge\  f_{i,k}\in F_i) \\
\end{split}
\end{equation*}
\end{fleqn}

We estimate metrics \emph{Fan-out} (\FO), \emph{Fan-in} (\FI)
and \emph{Module Instability} (\MI) for each module $m_i$,
on the basis of $F_{i, out}$, $F_{i, in}$:

\begin{align*}
\text{FO}_i & = |F_{i, out}|, \text{FI}_i = |F_{i, in}\ | \\
\text{I}_i  & = \frac{|F_{i, out}|}{|F_{i, out}| + |F_{i, in}|}
\end{align*}

\begingroup
\setlength{\tabcolsep}{6pt} 
\renewcommand{\arraystretch}{1.2} 
\begin{table*}[tb]
  \centering 
\label{tbl:module-coupling}
{
\scriptsize
\begin{tabular}{@{\extracolsep{2pt}}l r r r r r r@{}}
\toprule
\multirow{2}{2.0cm}{\textbf{Project}} &
\multicolumn{2}{c}{\textbf{Fan-out (\avgFO)}} &
\multirow{2}{1.8cm}{\textbf{Fan-in (\avgFI)}} &
\multicolumn{3}{c}{\textbf{Module Instability (\avgMI)}} \\

\cmidrule{2-3} \cmidrule{5-7}
  & \multicolumn{1}{c}{ES5} 
  & \multicolumn{1}{c}{ES6} 
  &
  & \multicolumn{1}{c}{ES5}
  & \multicolumn{1}{c}{ES6}
  & \multicolumn{1}{c}{\avgDI \%}\\ \hline

UltraTetris & 3.20 (0.45) & 0.80 (1.30) & 0.80 (0.45) & 0.80 (0.11) & 0.35 (0.49) & 60 (55) \\
Hangman & 21.33 (2.29) & 2.56 (3.57) & 9.11 (13.15) & 0.79 (0.25) & 0.37 (0.42) & 58 (44) \\
TicTacToe & 2.50 (3.54) & 1.50 (2.12) & 0.50 (0.71) & 0.50 (0.71) & 0.50 (0.71) &  0 ( 0) \\
uki & 10.00 (1.00) & \change{2.67 (3.79)} & 1.00 (1.00) & 0.91 (0.09) & 0.44 (0.51) & 53 (50) \\
\hline
GluttonousSnake & 1.75 (2.87) & 1.75 (2.87) & 0.75 (0.50) & 0.46 (0.54) & 0.46 (0.54) &  0 ( 0) \\
astix & 12.80 (13.61) & 10.40 (10.95) & 2.60 (3.10) & 0.71 (0.33) & 0.64 (0.44) & 25 (46) \\
game-of-life & 0.71 (1.11) & 0.71 (1.11) & 1.43 (1.62) & 0.32 (0.43) & 0.32 (0.43) &  0 ( 0) \\
tetrisJS & 10.27 (5.50) & 1.18 (2.04) & 1.91 (2.30) & 0.77 (0.31) & 0.37 (0.39) & 51 (42) \\
dynablaster-js-port & \change{3.27 (3.83)} & \change{2.02 (2.76)} & 2.73 (4.30) & \change{0.52 (0.39)} & \change{0.44 (0.37)} & 11 (19) \\
\hline
backbone-tableview & 2.08 (2.94) & 1.42 (2.31) & 1.08 (1.00) & 0.39 (0.41) & 0.39 (0.41) &  0 ( 0) \\
easystarjs & 0.67 (1.15) & 0.67 (1.15) & 0.67 (0.58) & 0.33 (0.58) & 0.33 (0.58) &  0 ( 0) \\
geojsonhint & 0.50 (0.58) & 0.50 (0.58) & 0.50 (0.58) & 0.50 (0.50) & 0.50 (0.50) &  0 ( 0) \\
session & 1.40 (1.67) & 1.00 (1.22) & 8.80 (6.38) & 0.24 (0.43) & 0.23 (0.43) &  9 (21) \\
underscore.string & 2.63 (9.84) & \change{1.82 (7.58)} & 1.02 (3.61) & 0.71 (0.34) & 0.70 (0.34) &  1 ( 7) \\
messy & 2.94 (4.95) & \change{2.69 (4.88)} & 6.31 (10.68) & 0.40 (0.43) & \change{0.39 (0.42)} &  \change{1 ( 3)} \\
virtual-dom & 2.30 (3.88) & 1.73 (3.01) & 1.45 (1.79) & 0.49 (0.43) & 0.48 (0.42) &  2 ( 5) \\
recipe-parser & 35.62 (23.87) & \change{1.06 (1.61)} & 13.88 (29.81) & \change{0.71 (0.40)} & \change{0.28 (0.38)} & \change{52 (44)} \\
planck.js & \change{70.95 (55.75)} & \change{10.22 (7.97)} & 24.42 (55.98) & 0.73 (0.39) & \change{0.60 (0.40)} & \change{20 (27)} \\
goojs & \change{26.40 (44.47)} & \change{16.20 (39.25)} & 10.74 (43.94) & 0.70 (0.37) & 0.63 (0.41) & 19 (34) \\

\bottomrule
\end{tabular}
}
\caption{Module coupling and instability metrics before and after refactoring.}
\label{Tab:moduleCouplingMetrics}
\end{table*}
\endgroup

Table~\ref{Tab:moduleCouplingMetrics} presents
module coupling and instability results for each project,
before and after refactoring.
Each value is computed as the average of a given metric
over all modules in the production code of the respective project.
The standard deviation is provided inside parentheses.
Columns 2--3 present the average \emph{Fan-out} (\avgFO),
for the modules of each project,
before (ES5) and after refactoring (ES6).
The results show 
a significant decrease of module \emph{Fan-out} in most projects,
marginal decrease in one project and
no change in {\color{black}5} of them.
The improvement of module \emph{Fan-out} is due
to the fine-grained reuse of module features after refactoring,
where each module imports 
the minimum required features from its dependencies.
On the other hand, the ES5 version of modules,
usually, import entire modules
and depend on all features bound to the respective module objects.
The average \emph{Fan-in} (\avgFI) of modules for each project
is presented in Column 4.
The values for \emph{Fan-in} are not affected by the refactoring.
The reason is that the incoming dependencies of a module remain unchanged,
despite its import in client modules
as individual module features 
rather than as a coarse-grained module object.

The effect of the refactoring on \emph{Module Instability}
is analyzed in Columns 5--6 that provide the average instability of modules
across all projects,
before and after refactoring,
Its average improvement is provided in Column 7. 
Following the decrease of module \emph{Fan-out},
the instability of modules improves
by \change{9--60\%} in 10 projects and
1--2\% in 3 projects.
However, in projects \codew{TicTacToe} and
\codew{backbone-tableview} 
module instability remains the same
despite the improvement in \emph{Fan-out}.
These projects contain modules with either
(a) \FI$=0$, i.e. their client modules 
do not define module features that 
depend on the exported features, or
(b) \FO$=0$, i.e. they do not define module features that 
depend on the imported features.
These modules are not considered in the calculation of
\avgMI after refactoring.

\changeStart
\begin{answer}
The proposed method increases
the number of reusable elements per refactored project
by a factor of 4 on average,
and enables fine-grained reuse
through declaration of multiple reusable elements per module.
It enforces module encapsulation
by restricting the scope of module features
that are not used by other modules
and reduces module coupling in terms of
decreased module Fan-out and module instability.
\end{answer}
\changeEnd

\subsubsection{RQ4: 
Does the proposed source code transformation preserve 
the external behaviour of the analyzed system?} 

To evaluate how effectively 
the proposed method preserves the external behaviour 
of each analyzed project, 
we executed the project's test suite 
after the proposed refactoring.
\changeStart
Its successful execution provides an indicator 
for the reliability of the applied refactoring \cite{Gligoric2013}.
\changeEnd
For projects without a test suite, 
we empirically evaluated the correctness of the refactored code
through manual code inspection and system usage.

Evaluation results provide empirical evidence
on the soundness of the code transformation,
since the test suites and the manual inspection
did not reveal changes to 
the external behaviour of refactored code.
Special attention needs to be attributed to the \goojs{} project,
where the transformation was applied despite the violation of a precondition. 
Specifically, \goojs{} modules contain module dependencies that 
are established in inner scopes (e.g. inside conditional statements). 
Therefore, their code violates the module format precondition 
with respect to nested module dependencies (\ref{sect:preconditions}). 
Nevertheless, these dependencies did not exhibit side-effects 
(e.g. modifications to the global scope)
and the establishment of these dependencies at the module scope
after refactoring did not alter the external behaviour of the system.

\begin{answer}
The proposed transformation preserves 
each analyzed system's external behaviour, even in case that 
it violates the precondition relevant to module dependencies established 
in inner scopes. 
\end{answer}

\changeStart
\section{Threats to validity}\label{threats}

In this section, 
we discuss threats to the validity of empirical evaluation results.
We categorize and prioritize the threats to validity 
according to~\citet{Wohlin2000}:
internal, external, construct and conclusion validity.

\emph{Internal validity} threats refer to factors
that affect the results of this study 
but are ignorred or cannot be controlled.
A potential threat relates to the quality 
of the benchmark projects' test suites
that affects the conclusions of RQ4.
Since the statement coverage of test suites
in CJS projects ranges between \pct{40.05}--\pct{100} 
and in 7 of them exceeds \pct{90}
(Table~\ref{Tab:projectImplementationDetails}),
we expect that they are capable of
revealing the majority of flaws of the proposed  method.

\emph{External validity} threats affect 
the generalization of empirical evaluation results.
A potential threat concerns the extent at which
the selected sample of 19 projects
is representative of the population of legacy ES5 projects.
We deem that the sample has adequate size and diversity
in order to mitigate this threat,
since it includes both standalone applications and library projects,
with various sizes and coming from different domains.
The general applicability of the proposed method is,
also, affected by some limitations on handling specific 
features of the JavaScript language.
Specifically, our method does not analyze code
which uses dynamic code generation and name aliasing, 
e.g. resolving modules with variant file paths generated
using \texttt{\_\_filename} or \texttt{\_\_dirname}, 
properties defined programmatically
through the \codew{Object.defineProperty()} method.
Static analysis introduces significant challenges
in these cases~\cite{Jensen12} 
that may undermine the safety of the proposed refactoring.
Finally, the method analyzes to a limited extent 
language features that are forbidden in strict mode~\cite{mdnJSRef},
in order to prevent the introduction of defects
due to incompatibilities between the ES5 and ES6 specifications.
Among benchmark projects,
\GluttonousSnake{}, \astix{}, \gameOfLife{}, 
\UltraTetris{}, \tetrisjs{} and \texttt{goojs}
make limited use of such features.
In order to include them in this study,
we manually replace them with equivalent features, 
permitted in strict mode.

Threats to \emph{construct validity} concern the correspondence 
between theoretical constructs 
and the observations of the empirical evaluation.
A potential threat relates to RQ1 and the regex classifiers
that we employed in order to categorize JS files to module formats.
In order to mitigate this threat we built a test suite
for these regular expressions that is publicly available in Github
\footnote{https://github.com/softeng-aueb/es-module-classifier}.
Another threat relates to the correct estimation of metrics
that support the conclusions of RQ2 and RQ3.
In order to contain this threat
with performed manual validation of metric calculations
for small projects that are included in this study.

\emph{Conclusion validity} concerns the statistical significance
of the results on cause-effect relations studied in the experiment. 
Since, we conduct an exploratory study, 
where experimental results and observations 
shape the findings of the study rather than confirm the validity of certain hypotheses, 
we evaluate \emph{reliability} as a counterpart of conclusion validity~\cite{Wohlin2000}.
Threats to \emph{reliability} concern the reproducibility of this study
and we mitigate them through publishing the source code
and all required artifacts for its replication.  
\changeEnd
\section{Results and Discussion}
\label{results}

The proposed method automates a large refactoring 
that migrates an ES5 codebase to ECMAScript 6.
The refactoring focuses on the application
and effective use of the ES6 language constructs for modularity.
This migration represents an important milestone
to the evolution of a legacy JavaScript project,
since the use of the ES6 module system
is a basic requirement for further adoption
of language features (e.g. class syntax, promises API)
introduced in ES6 or beyond,
that boost developer productivity and improve code maintainability.

The scope of the proposed refactoring method
is not limited to plain syntax transformation,
that is supported to an extent by some module bundlers~\cite{rollupjs}.
It additionally alters the design of refactored modules 
to enable fine-grained reuse of module contents,
through destructuring of module objects
to multiple individually reusable module features.
Module dependencies are, then, redefined
and established on the basis of these module features.  
This improves maintainability, since 
coupling between the refactored modules is reduced; 
each refactored module imports 
the module features that are used by its features, 
instead of entire module objects. 
It, also, enables encapsulation,
since unused module features are restricted to 
the scope of their declaration module.

Our method achieves fine-grained reuse of module contents
through effective use of the \emph{named} imports/exports
language constructs,
instead of their \emph{default} or \emph{namespace} equivalents
that establish dependencies
on the basis of coarse-grained module objects.
Although the latter are used 
for backwards compatibility with ES5 module formats,
named imports/exports provide significant benefits
\footnote{https://esdiscuss.org/topic/moduleimport}.
Specifically, they enforce name consistency across modules,
which improves code understandability
and allows for effective ``tree-shaking'' during code deployment. 
Moreover, they facilitate other refactoring operations,
such as rename refactorings and move feature
or extract module refactorings for remodularization.
Finally, the evaluation part of this study
contributes to the ongoing debate 
on \emph{named} vs. \emph{default} imports/exports\footnotemark{}
by highlighting the former's positive effects 
on module coupling and reusability. 
\footnotetext{N.Zakas, https://humanwhocodes.com/blog/2019/01/stop-using-default-exports-javascript-module/, 
A. Rauschmayer, https://medium.com/@rauschma/note-that-default-exporting-objects-is-usually-an-anti-pattern-if-you-want-to-export-the-cf674423ac38}

\section{Conclusions -- Future Work} \label{Conclusions}

We have proposed a method 
for automated refactoring of legacy ES5 code,
that is either non-modular or uses the AMD/CJS formats,
to ES6 modularity with extensive use
of the \emph{named import/export} language constructs.
Besides migration of ES5 module formats to ES6 modules,
the proposed method improves internal quality attributes
of refactored code.
Specifically, it enforces fine-grained reuse of module contents,
through destructuring of module objects
to individually reusable module features.
On the basis of these features, 
it optimizes module dependencies by leveraging the ES6 syntax.
The core of our method
is a static analysis technique
that constructs a global model of the analyzed project, 
the Module Dependence Graph (MDG). 
The MDG mirrors inferred or explicitly declared modules
and their dependencies.
On the basis of MDG, we specify the source code transformation
for migration to ES6. 

\changeStart
The proposed method has been implemented in the Node.js platform
for the purpose of evaluating its effectiveness and practicality
on a collection of open source projects.
\changeEnd 
Evaluation results reveal a developer intent for fine-grained reuse,
since \change{\pct{78.6}} of extracted features have semantics 
that correspond to reusable elements at module scope
(namespace properties, static features of emulated classes).
The analysis of refactored code shows
an increase in the number of reusable elements per project by a factor of four
and an improvement in the coupling of refactored modules
in terms of decreased module \emph{Fan-out} and module instability.
Finally, the execution of projects’
test-suites on refactored code provides evidence on the
soundness of the source code transformation.

Our future work will investigate the potential
for further improvements to the modularity of an ES6 codebase
through automated refactorings.
Specifically, we intend to extend our research  
across two directions:
(a) improvement of module cohesion through redistribution 
of fine-grained module features among modules,
extraction of module features to new modules or 
consolidation of the contents of different modules,
and (b) modernization of the internal structure of modules
through the introduction of the ES6 class syntax.

\changeStart
\section*{Acknowledgements}

The research work was supported by 
the Hellenic Foundation for
Research and Innovation (HFRI) under 
the HFRI PhD Fellowship grant
(Fellowship Number: 696).
The authors would, also, like to thank the anonymous reviewers
for their useful comments that improved the quality of this work.
\changeEnd


\begin{thebibliography}{50}
\expandafter\ifx\csname natexlab\endcsname\relax\def\natexlab#1{#1}\fi
\providecommand{\url}[1]{\texttt{#1}}
\providecommand{\href}[2]{#2}
\providecommand{\path}[1]{#1}
\providecommand{\DOIprefix}{doi:}
\providecommand{\ArXivprefix}{arXiv:}
\providecommand{\URLprefix}{URL: }
\providecommand{\Pubmedprefix}{pmid:}
\providecommand{\doi}[1]{\href{http://dx.doi.org/#1}{\path{#1}}}
\providecommand{\Pubmed}[1]{\href{pmid:#1}{\path{#1}}}
\providecommand{\bibinfo}[2]{#2}
\ifx\xfnm\relax \def\xfnm[#1]{\unskip,\space#1}\fi
\bibitem[{Wirfs-Brock and Eich(2020)}]{Wirfs2020}
\bibinfo{author}{A.~Wirfs-Brock}, \bibinfo{author}{B.~Eich},
\newblock \bibinfo{title}{Javascript: The first 20 years},
\newblock \bibinfo{journal}{Proc. ACM Program. Lang.} \bibinfo{volume}{4}
  (\bibinfo{year}{2020}). \DOIprefix\doi{10.1145/3386327}.
\bibitem[{es6(2015)}]{es6spec}
\bibinfo{title}{{ECMAScript 2015 Language Specification}},
  \bibinfo{howpublished}{https://www.ecma-international.org/ecma-262/ 6.0/},
  \bibinfo{year}{June 2015}.
\bibitem[{{MDN Web Docs}(2020)}]{mdnJSRef}
\bibinfo{author}{{MDN Web Docs}}, \bibinfo{title}{Javascript reference},
  \bibinfo{howpublished}{https://developer.mozilla.org/en-US/docs/Web/
  JavaScript/Reference}, \bibinfo{year}{2020}.
\bibitem[{nod(2020)}]{nodejsEsm}
\bibinfo{title}{{Node.js v12.6.0 Documentation: ECMAScript Modules}},
  \bibinfo{howpublished}{https://nodejs.org/api/esm.html},
  \bibinfo{year}{Accessed: July 2020}.
\bibitem[{amd(2020)}]{amdspec}
\bibinfo{title}{{Asynchronous Module Definition (AMD) API}},
  \bibinfo{howpublished}{https://github.com/amdjs/amdjs-api},
  \bibinfo{year}{Accessed: July 2020}.
\bibitem[{nod(2019)}]{nodejsModules}
\bibinfo{title}{{Node.js Documentation: CommonJS Modules}},
  \bibinfo{howpublished}{https://nodejs.org/api/modules.html},
  \bibinfo{year}{2019}.
\bibitem[{Rauschmayer(2015)}]{Rauschmayer15}
\bibinfo{author}{A.~Rauschmayer}, \bibinfo{title}{Exploring ES6. Upgrade to the
  next version of JavaScript}, \bibinfo{year}{2015}.
\bibitem[{Shore(2004)}]{Shore2004}
\bibinfo{author}{J.~Shore},
\newblock \bibinfo{title}{Fail fast},
\newblock \bibinfo{journal}{IEEE Software} \bibinfo{volume}{21}
  (\bibinfo{year}{2004}) \bibinfo{pages}{21--25}.
  \DOIprefix\doi{10.1109/MS.2004.1331296}.
\bibitem[{Mens and Tourw{\'e}(2004)}]{Mens2004}
\bibinfo{author}{T.~Mens}, \bibinfo{author}{T.~Tourw{\'e}},
\newblock \bibinfo{title}{A survey of software refactoring},
\newblock \bibinfo{journal}{IEEE Transactions on software engineering}
  \bibinfo{volume}{30} (\bibinfo{year}{2004}) \bibinfo{pages}{126--139}.
\bibitem[{Fowler(1999)}]{Fowler99}
\bibinfo{author}{M.~Fowler}, \bibinfo{title}{Refactoring: Improving the Design
  of Existing Code}, \bibinfo{edition}{1st} ed.,
  \bibinfo{publisher}{Addison-Wesley Longman Publishing Co., Inc.},
  \bibinfo{address}{Boston, MA, USA}, \bibinfo{year}{1999}.
\bibitem[{Fowler(2018)}]{Fowler18}
\bibinfo{author}{M.~Fowler}, \bibinfo{title}{Refactoring: Improving the Design
  of Existing Code}, \bibinfo{edition}{2nd} ed.,
  \bibinfo{publisher}{Addison-Wesley Longman Publishing Co., Inc.},
  \bibinfo{address}{Boston, MA, USA}, \bibinfo{year}{2018}.
\bibitem[{Sousa et~al.(2020)Sousa, Oizumi, Garcia, Oliveira, Cedrim, and
  Lucena}]{Sousa2020}
\bibinfo{author}{L.~Sousa}, \bibinfo{author}{W.~Oizumi},
  \bibinfo{author}{A.~Garcia}, \bibinfo{author}{A.~Oliveira},
  \bibinfo{author}{D.~Cedrim}, \bibinfo{author}{C.~Lucena},
\newblock \bibinfo{title}{When are smells indicators of architectural
  refactoring opportunities: A study of 50 software projects},
\newblock in: \bibinfo{booktitle}{Proceedings of the 28th International
  Conference on Program Comprehension}, \bibinfo{year}{2020}, pp.
  \bibinfo{pages}{354--365}.
\bibitem[{Soares et~al.(2020)Soares, Ribeiro, Amaral, Gheyi, Fernandes, Garcia,
  Fonseca, and Santos}]{Soares2020}
\bibinfo{author}{E.~Soares}, \bibinfo{author}{M.~Ribeiro},
  \bibinfo{author}{G.~Amaral}, \bibinfo{author}{R.~Gheyi},
  \bibinfo{author}{L.~Fernandes}, \bibinfo{author}{A.~Garcia},
  \bibinfo{author}{B.~Fonseca}, \bibinfo{author}{A.~Santos},
\newblock \bibinfo{title}{Refactoring test smells: A perspective from
  open-source developers},
\newblock in: \bibinfo{booktitle}{Proceedings of the 5th Brazilian Symposium on
  Systematic and Automated Software Testing}, \bibinfo{year}{2020}, pp.
  \bibinfo{pages}{50--59}.
\bibitem[{Bavota et~al.(2015)Bavota, De~Lucia, Di~Penta, Oliveto, and
  Palomba}]{Bavota2015}
\bibinfo{author}{G.~Bavota}, \bibinfo{author}{A.~De~Lucia},
  \bibinfo{author}{M.~Di~Penta}, \bibinfo{author}{R.~Oliveto},
  \bibinfo{author}{F.~Palomba},
\newblock \bibinfo{title}{An experimental investigation on the innate
  relationship between quality and refactoring},
\newblock \bibinfo{journal}{Journal of Systems and Software}
  \bibinfo{volume}{107} (\bibinfo{year}{2015}) \bibinfo{pages}{1--14}.
\bibitem[{Lacerda et~al.(2020)Lacerda, Petrillo, Pimenta, and
  Guéhéneuc}]{Lacerda2020}
\bibinfo{author}{G.~Lacerda}, \bibinfo{author}{F.~Petrillo},
  \bibinfo{author}{M.~Pimenta}, \bibinfo{author}{Y.~G. Guéhéneuc},
\newblock \bibinfo{title}{Code smells and refactoring: A tertiary systematic
  review of challenges and observations},
\newblock \bibinfo{journal}{Journal of Systems and Software}
  \bibinfo{volume}{167} (\bibinfo{year}{2020}).
\bibitem[{Fernandes et~al.(2020)Fernandes, Ch{\'a}vez, Garcia, Ferreira,
  Cedrim, Sousa, and Oizumi}]{Fernandes2020}
\bibinfo{author}{E.~Fernandes}, \bibinfo{author}{A.~Ch{\'a}vez},
  \bibinfo{author}{A.~Garcia}, \bibinfo{author}{I.~Ferreira},
  \bibinfo{author}{D.~Cedrim}, \bibinfo{author}{L.~Sousa},
  \bibinfo{author}{W.~Oizumi},
\newblock \bibinfo{title}{Refactoring effect on internal quality attributes:
  What haven’t they told you yet?},
\newblock \bibinfo{journal}{Information and Software Technology}
  (\bibinfo{year}{2020}) \bibinfo{pages}{106347}.
\bibitem[{Agnihotri and Chug(2020)}]{Agnihotri2020}
\bibinfo{author}{M.~Agnihotri}, \bibinfo{author}{A.~Chug},
\newblock \bibinfo{title}{A systematic literature survey of software metrics,
  code smells and refactoring techniques},
\newblock \bibinfo{journal}{Journal of Information Processing Systems}
  \bibinfo{volume}{16} (\bibinfo{year}{2020}) \bibinfo{pages}{915--934}.
\bibitem[{Baqais and Alshayeb(2020)}]{Baqais2020}
\bibinfo{author}{A.~A.~B. Baqais}, \bibinfo{author}{M.~Alshayeb},
\newblock \bibinfo{title}{Automatic software refactoring: a systematic
  literature review},
\newblock \bibinfo{journal}{Software Quality Journal} \bibinfo{volume}{28}
  (\bibinfo{year}{2020}) \bibinfo{pages}{459--502}.
\bibitem[{Abid et~al.(2020)Abid, Alizadeh, Kessentini, do~Nascimento~Ferreira,
  and Dig}]{Abid2020}
\bibinfo{author}{C.~Abid}, \bibinfo{author}{V.~Alizadeh},
  \bibinfo{author}{M.~Kessentini}, \bibinfo{author}{T.~do~Nascimento~Ferreira},
  \bibinfo{author}{D.~Dig}, \bibinfo{title}{30 years of software refactoring
  research:a systematic literature review}, \bibinfo{year}{2020}.
  \href{http://arxiv.org/abs/2007.02194}{{\tt arXiv:2007.02194}}.
\bibitem[{Nasagh et~al.(2020)Nasagh, Shahidi, and Ashtiani}]{Nasagh2020}
\bibinfo{author}{R.~S. Nasagh}, \bibinfo{author}{M.~Shahidi},
  \bibinfo{author}{M.~Ashtiani},
\newblock \bibinfo{title}{A fuzzy genetic automatic refactoring approach to
  improve software maintainability and flexibility},
\newblock \bibinfo{journal}{Soft Computing}  (\bibinfo{year}{2020})
  \bibinfo{pages}{1--31}.
\bibitem[{Bruce et~al.(2020)Bruce, Zhang, Arora, Xu, and Kim}]{Bruce2020}
\bibinfo{author}{B.~R. Bruce}, \bibinfo{author}{T.~Zhang},
  \bibinfo{author}{J.~Arora}, \bibinfo{author}{G.~H. Xu},
  \bibinfo{author}{M.~Kim},
\newblock \bibinfo{title}{Jshrink: In-depth investigation into debloating
  modern java applications},
\newblock in: \bibinfo{booktitle}{Proceedings of the 28th ACM Joint Meeting on
  European Software Engineering Conference and Symposium on the Foundations of
  Software Engineering}, ESEC/FSE 2020, \bibinfo{publisher}{ACM},
  \bibinfo{year}{2020}, p. \bibinfo{pages}{135–146}.
\bibitem[{de~Freitas~Brito et~al.(2020)de~Freitas~Brito, Hora, and
  Valente}]{Brito2020}
\bibinfo{author}{A.~de~Freitas~Brito}, \bibinfo{author}{A.~C. Hora},
  \bibinfo{author}{M.~T. Valente},
\newblock \bibinfo{title}{Refactoring graphs: Assessing refactoring over time},
\newblock \bibinfo{journal}{2020 IEEE 27th International Conference on Software
  Analysis, Evolution and Reengineering (SANER)}  (\bibinfo{year}{2020})
  \bibinfo{pages}{367--377}.
\bibitem[{Paiva et~al.(2020)Paiva, Freire, and de~Mattos~Fortes}]{Paiva2020}
\bibinfo{author}{D.~M.~B. Paiva}, \bibinfo{author}{A.~P. Freire},
  \bibinfo{author}{R.~P. de~Mattos~Fortes},
\newblock \bibinfo{title}{Accessibility and software engineering processes: A
  systematic literature review},
\newblock \bibinfo{journal}{Journal of Systems and Software}
  (\bibinfo{year}{2020}) \bibinfo{pages}{110819}.
\bibitem[{Ivers et~al.(2020)Ivers, Ozkaya, Nord, and Seifried}]{Ivers2020}
\bibinfo{author}{J.~Ivers}, \bibinfo{author}{I.~Ozkaya}, \bibinfo{author}{R.~L.
  Nord}, \bibinfo{author}{C.~Seifried},
\newblock \bibinfo{title}{Next generation automated software evolution
  refactoring at scale},
\newblock in: \bibinfo{booktitle}{Proceedings of the 28th ACM Joint Meeting on
  European Software Engineering Conference and Symposium on the Foundations of
  Software Engineering}, \bibinfo{year}{2020}, pp. \bibinfo{pages}{1521--1524}.
\bibitem[{Ying and Miller(2013)}]{Ying2013}
\bibinfo{author}{M.~Ying}, \bibinfo{author}{J.~Miller},
\newblock \bibinfo{title}{Refactoring legacy ajax applications to improve the
  efficiency of the data exchange component},
\newblock \bibinfo{journal}{Journal of Systems and Software}
  \bibinfo{volume}{86} (\bibinfo{year}{2013}) \bibinfo{pages}{72–88}.
\bibitem[{Obbink et~al.(2018)Obbink, Malavolta, Scoccia, and Lago}]{Obbink2018}
\bibinfo{author}{N.~G. Obbink}, \bibinfo{author}{I.~Malavolta},
  \bibinfo{author}{G.~L. Scoccia}, \bibinfo{author}{P.~Lago},
\newblock \bibinfo{title}{An extensible approach for taming the challenges of
  javascript dead code elimination},
\newblock in: \bibinfo{booktitle}{2018 IEEE 25th International Conference on
  Software Analysis, Evolution and Reengineering (SANER)},
  \bibinfo{year}{2018}, pp. \bibinfo{pages}{291--401}.
\bibitem[{Vazquez et~al.(2018)Vazquez, Bergel, Vidal, Pace, and
  Marcos}]{Vazquez2018}
\bibinfo{author}{H.~Vazquez}, \bibinfo{author}{A.~Bergel},
  \bibinfo{author}{S.~Vidal}, \bibinfo{author}{J.~Pace},
  \bibinfo{author}{C.~Marcos},
\newblock \bibinfo{title}{Slimming javascript applications: an approach for
  removing unused functions from javascript libraries},
\newblock \bibinfo{journal}{Information and Software Technology}
  (\bibinfo{year}{2018}).
\bibitem[{Feldthaus et~al.(2011)Feldthaus, Millstein, M{\o}ller, Sch\"{a}fer,
  and Tip}]{Feldthaus11}
\bibinfo{author}{A.~Feldthaus}, \bibinfo{author}{T.~Millstein},
  \bibinfo{author}{A.~M{\o}ller}, \bibinfo{author}{M.~Sch\"{a}fer},
  \bibinfo{author}{F.~Tip},
\newblock \bibinfo{title}{Tool-supported refactoring for javascript},
\newblock in: \bibinfo{booktitle}{Proceedings of the 2011 ACM Int. Conf. on
  Object Oriented Programming Systems Languages and Applications}, OOPSLA '11,
  \bibinfo{publisher}{ACM}, \bibinfo{year}{2011}, pp.
  \bibinfo{pages}{119--138}.
\bibitem[{Feldthaus and M{\o}ller(2013)}]{Feldthaus13}
\bibinfo{author}{A.~Feldthaus}, \bibinfo{author}{A.~M{\o}ller},
\newblock \bibinfo{title}{Semi-automatic rename refactoring for javascript},
\newblock \bibinfo{journal}{SIGPLAN Not.} \bibinfo{volume}{48}
  (\bibinfo{year}{2013}) \bibinfo{pages}{323--338}.
\bibitem[{Jensen et~al.(2012)Jensen, Jonsson, and M{\o}ller}]{Jensen12}
\bibinfo{author}{S.~H. Jensen}, \bibinfo{author}{P.~A. Jonsson},
  \bibinfo{author}{A.~M{\o}ller},
\newblock \bibinfo{title}{Remedying the eval that men do},
\newblock in: \bibinfo{booktitle}{Proceedings of the 2012 Int. Symp. on
  Software Testing and Analysis}, ISSTA 2012, \bibinfo{publisher}{ACM},
  \bibinfo{year}{2012}, pp. \bibinfo{pages}{34--44}.
\bibitem[{{Gallaba} et~al.(2015){Gallaba}, {Mesbah}, and
  {Beschastnikh}}]{Gallaba2015}
\bibinfo{author}{K.~{Gallaba}}, \bibinfo{author}{A.~{Mesbah}},
  \bibinfo{author}{I.~{Beschastnikh}},
\newblock \bibinfo{title}{Don't call us, we'll call you: Characterizing
  callbacks in javascript},
\newblock in: \bibinfo{booktitle}{2015 ACM/IEEE Int. Symp. on Empirical
  Software Engineering and Measurement (ESEM)}, \bibinfo{year}{2015}.
\bibitem[{Brodu et~al.(2015)Brodu, Fr{\'e}not, and Obl{\'e}}]{Brodu2015}
\bibinfo{author}{E.~Brodu}, \bibinfo{author}{S.~Fr{\'e}not},
  \bibinfo{author}{F.~Obl{\'e}},
\newblock \bibinfo{title}{Toward automatic update from callbacks to promises},
\newblock in: \bibinfo{booktitle}{Proceedings of the 1st Workshop on All-Web
  Real-Time Systems}, AWeS '15, \bibinfo{publisher}{ACM}, \bibinfo{year}{2015}.
\bibitem[{{Gallaba} et~al.(2017){Gallaba}, {Hanam}, {Mesbah}, and
  {Beschastnikh}}]{Gallaba2017}
\bibinfo{author}{K.~{Gallaba}}, \bibinfo{author}{Q.~{Hanam}},
  \bibinfo{author}{A.~{Mesbah}}, \bibinfo{author}{I.~{Beschastnikh}},
\newblock \bibinfo{title}{Refactoring asynchrony in javascript},
\newblock in: \bibinfo{booktitle}{2017 IEEE Int. Conf. on Software Maintenance
  and Evolution (ICSME)}, \bibinfo{year}{2017}, pp. \bibinfo{pages}{353--363}.
\bibitem[{Rostami et~al.(2016)Rostami, Eshkevari, Mazinanian, and
  Tsantalis}]{Rostami16}
\bibinfo{author}{S.~Rostami}, \bibinfo{author}{L.~Eshkevari},
  \bibinfo{author}{D.~Mazinanian}, \bibinfo{author}{N.~Tsantalis},
\newblock \bibinfo{title}{{Detecting Function Constructors in JavaScript}},
\newblock in: \bibinfo{booktitle}{2016 IEEE Int. Conf. on Software Maintenance
  and Evolution (ICSME)}, \bibinfo{year}{2016}, pp. \bibinfo{pages}{488--492}.
\bibitem[{Silva et~al.(2017{\natexlab{a}})Silva, Valente, and
  Bergel}]{SilvaSANER17}
\bibinfo{author}{L.~H. Silva}, \bibinfo{author}{M.~T. Valente},
  \bibinfo{author}{A.~Bergel},
\newblock \bibinfo{title}{Statically identifying class dependencies in legacy
  javascript systems: First results},
\newblock in: \bibinfo{booktitle}{2017 IEEE 24th Int. Conf. on Software
  Analysis, Evolution and Reengineering (SANER)},
  \bibinfo{year}{2017}{\natexlab{a}}.
\bibitem[{Silva et~al.(2017{\natexlab{b}})Silva, Valente, Bergel, Anquetil, and
  Etien}]{SilvaJSEP17}
\bibinfo{author}{L.~H. Silva}, \bibinfo{author}{M.~T. Valente},
  \bibinfo{author}{A.~Bergel}, \bibinfo{author}{N.~Anquetil},
  \bibinfo{author}{A.~Etien},
\newblock \bibinfo{title}{Identifying classes in legacy {JavaScript} code},
\newblock \bibinfo{journal}{Journal of Software: Evolution and Process}
  \bibinfo{volume}{1} (\bibinfo{year}{2017}{\natexlab{b}})
  \bibinfo{pages}{1--37}.
\bibitem[{Silva et~al.(2017{\natexlab{c}})Silva, Valente, and
  Bergel}]{SilvaICSR17}
\bibinfo{author}{L.~H. Silva}, \bibinfo{author}{M.~T. Valente},
  \bibinfo{author}{A.~Bergel},
\newblock \bibinfo{title}{Refactoring legacy {JavaScript} code to use classes:
  The good, the bad and the ugly},
\newblock in: \bibinfo{booktitle}{16th Int. Conf. on Software Reuse (ICSR)},
  \bibinfo{year}{2017}{\natexlab{c}}, pp. \bibinfo{pages}{1--16}.
\bibitem[{Paltoglou et~al.(2018)Paltoglou, Zafeiris, Giakoumakis, and
  Diamantidis}]{Paltoglou2018}
\bibinfo{author}{A.~Paltoglou}, \bibinfo{author}{V.~E. Zafeiris},
  \bibinfo{author}{E.~A. Giakoumakis}, \bibinfo{author}{N.~Diamantidis},
\newblock \bibinfo{title}{Automated refactoring of client-side javascript code
  to es6 modules},
\newblock in: \bibinfo{booktitle}{2018 IEEE 25th Int. Conf. on Software
  Analysis, Evolution and Reengineering (SANER)}, \bibinfo{organization}{IEEE},
  \bibinfo{year}{2018}, pp. \bibinfo{pages}{402--412}.
\bibitem[{Rauschmayer(2014)}]{Rauschmayer2014}
\bibinfo{author}{A.~Rauschmayer}, \bibinfo{title}{Speaking JavaScript},
  \bibinfo{edition}{1st} ed., \bibinfo{publisher}{O'Reilly Media, Inc.},
  \bibinfo{year}{2014}.
\bibitem[{Osmani(2012)}]{Osmani08}
\bibinfo{author}{A.~Osmani}, \bibinfo{title}{Learning JavaScript Design
  Patterns}, \bibinfo{publisher}{O'Reilly Media, Inc.}, \bibinfo{year}{2012}.
\bibitem[{Zakas(2012)}]{Zakas12}
\bibinfo{author}{N.~Zakas}, \bibinfo{title}{Maintainable JavaScript},
  \bibinfo{publisher}{O'Reilly Media, Inc.}, \bibinfo{year}{2012}.
\bibitem[{Crockford(2008)}]{Crockford08}
\bibinfo{author}{D.~Crockford}, \bibinfo{title}{JavaScript: The Good Parts},
  \bibinfo{publisher}{O'Reilly Media, Inc.}, \bibinfo{year}{2008}.
\bibitem[{umd(2020)}]{umdspec}
\bibinfo{title}{{Universal Module Definition (UMD) API}},
  \bibinfo{howpublished}{https://github.com/umdjs/umd},
  \bibinfo{year}{Accessed: July 2020}.
\bibitem[{req(2020)}]{requirejs}
\bibinfo{title}{{RequireJS module loader}},
  \bibinfo{howpublished}{https://requirejs.org}, \bibinfo{year}{Accessed: July
  2020}.
\bibitem[{Grover and Kunduru(2017)}]{Grover2017}
\bibinfo{author}{D.~Grover}, \bibinfo{author}{H.~P. Kunduru},
  \bibinfo{title}{ES6 for Humans: The Latest Standard of JavaScript ES2015 and
  Beyond}, \bibinfo{edition}{1st} ed., \bibinfo{publisher}{Apress},
  \bibinfo{address}{USA}, \bibinfo{year}{2017}.
\bibitem[{rol(2020)}]{rollupjs}
\bibinfo{title}{{Rollup module bundler}},
  \bibinfo{howpublished}{https://rollupjs.org}, \bibinfo{year}{Accessed: July
  2020}.
\bibitem[{web(2019)}]{webpackTreeShaking}
\bibinfo{title}{{Webpack module bundler. Tree Shaking}},
  \bibinfo{howpublished}{https://webpack.js.org /guides/tree-shaking/},
  \bibinfo{year}{Accessed: Nov. 2019}.
\bibitem[{Martin(2003)}]{Martin:2003}
\bibinfo{author}{R.~C. Martin}, \bibinfo{title}{Agile Software Development:
  Principles, Patterns, and Practices}, \bibinfo{publisher}{Prentice Hall PTR},
  \bibinfo{address}{Upper Saddle River, NJ, USA}, \bibinfo{year}{2003}.
\bibitem[{Gligoric et~al.(2013)Gligoric, Behrang, Li, Overbey, Hafiz, and
  Marinov}]{Gligoric2013}
\bibinfo{author}{M.~Gligoric}, \bibinfo{author}{F.~Behrang},
  \bibinfo{author}{Y.~Li}, \bibinfo{author}{J.~Overbey},
  \bibinfo{author}{M.~Hafiz}, \bibinfo{author}{D.~Marinov},
\newblock \bibinfo{title}{Systematic testing of refactoring engines on real
  software projects},
\newblock in: \bibinfo{booktitle}{Proceedings of the 27th European Conf. on
  Object-Oriented Programming}, ECOOP'13, \bibinfo{publisher}{Springer-Verlag},
  \bibinfo{year}{2013}, pp. \bibinfo{pages}{629--653}.
\bibitem[{Wohlin et~al.(2000)Wohlin, Runeson, H\"{o}st, Ohlsson, Regnell, and
  Wessl{\'e}n}]{Wohlin2000}
\bibinfo{author}{C.~Wohlin}, \bibinfo{author}{P.~Runeson},
  \bibinfo{author}{M.~H\"{o}st}, \bibinfo{author}{M.~C. Ohlsson},
  \bibinfo{author}{B.~Regnell}, \bibinfo{author}{A.~Wessl{\'e}n},
  \bibinfo{title}{Experimentation in Software Engineering: An Introduction},
  \bibinfo{publisher}{Kluwer Academic Publishers}, \bibinfo{address}{Norwell,
  MA, USA}, \bibinfo{year}{2000}.

\end{thebibliography}

\end{document}